\documentclass[a4]{aa}
\usepackage{graphicx}
\newcommand{\arcm}{\:{\rm arcmin}}
\newcommand{\arcs}{\:{\rm arcsec}}
\newcommand{\bmodel}{$\beta$--model\ }
\newcommand{\cm}{\:{\rm cm}}
\newcommand{\counts}{\:{\rm counts}}
\newcommand{\crocho}{\left[\begin{array}{cc}}
\newcommand{\crochf}{\end{array}\right]}
\newcommand{\fb}{f_{\rm b}}
\newcommand{\fg}{f_{\rm gas}}
\newcommand{\h}{\:h_{50}}
\newcommand{\kev}{\:{\rm keV}}
\newcommand{\km}{\:{\rm km}}
\newcommand{\lb}{L_{\rm B}}
\newcommand{\lbsun}{L_{{\rm B}\odot}}
\newcommand{\llim}{L_{\rm lim}}
\newcommand{\ltot}{L_{\rm tot}}
\newcommand{\lnep}{{\rm ln}}
\newcommand{\mblim}{M_{\rm B\,lim}}
\newcommand{\mbst}{M_{\rm B}^*}
\newcommand{\mvst}{M_{\rm V}^*}
\newcommand{\mbapp}{m_{\rm B}}
\newcommand{\mlb}{M/L_{\rm B}}
\newcommand{\mprot}{m_{\rm p}}
\newcommand{\msun}{M_{\odot}}
\newcommand{\mst}{M_*}
\newcommand{\mgas}{M_{\rm gas}}
\newcommand{\mtot}{M_{\rm tot}}
\newcommand{\mpc}{\:{\rm Mpc}}
\newcommand{\nez}{n_{\rm e0}}
\newcommand{\omz}{\Omega_0}
\newcommand{\omb}{\Omega_{\rm b}}
\newcommand{\paro}{\left(\begin{array}{cc}\!\!}
\newcommand{\parf}{\!\!\end{array}\right)}
\newcommand{\rc}{r_{\rm c}}
\newcommand{\rv}{r_{\rm v}}
\newcommand{\rcx}{r_{\rm cX}}
\newcommand{\rxl}{R_{\rm X\,lim}}
\newcommand{\s}{\:{\rm s}}
\newcommand{\thetac}{\theta_{\rm c}}
\newcommand{\thetamax}{\theta_{\rm max}}
\newcommand{\tx}{T_{\rm X}}
\newcommand{\txmv}{$T_{\rm X}\:$--$\:M_{\rm V}$\ }

\begin{document}
\thesaurus{02 (12.03.3;     %Cosmology:observations
               11.03.1;     %Galaxies:clusters
               13.25.3)     %X-rays: general
          }
\title{The Baryon Content of
Groups and Clusters of Galaxies}
\author{H. Roussel\inst{1}, R. Sadat\inst{2}, A. Blanchard\inst{2,3}}

\institute{ DAPNIA/Service d'Astrophysique, CEA/Saclay, 91191 Gif-sur-Yvette 
cedex, France  
\and 
Observatoire Midi-Pyr\'en\'ees, LAT, CNRS, 14 Av. Edouard Belin, 31~400, Toulouse, France 
\and 
Observatoire Astronomique, 
11, rue de l'Universit\'e, 67000 Strasbourg, France 
          }

\offprints{R. Sadat}

\date{Received \rule{2.0cm}{0.01cm} ; accepted \rule{2.0cm}{0.01cm} }

\titlerunning{Baryon Content of Groups and Clusters}
\maketitle

%***************************************

\begin{abstract}
We have analyzed the properties of a sample of 33 groups and clusters of galaxies for which both optical and X-ray data were available in the literature. This sample was built to examine the baryon content and to check for trends over a decade in temperature down to 1 keV.\\
We examine the relative contribution of galaxies and ICM to baryons in clusters through the gas-to-stellar mass ratio ($M_{gas}/M_{*}$). We find that the typical stellar contribution to the baryonic mass is between 5 and 20\%, at the virial radius. The ratio ($M_{gas}/M_{*}$) is found to be roughly independent of temperature. Therefore, we do not confirm the trend of increasing gas-to-stellar mass ratio with increasing temperature as previously claimed.\\
We also determine the absolute values and the distribution of the baryon fraction with the density contrast $\delta$ with respect to the critical density. Virial masses are estimated from two different mass estimators: one based on the isothermal hydrostatic equation (IHE), the other based on scaling law models (SLM), the calibration being taken from numerical simulations. 
Comparing the two methods, we find that SLM lead to less dispersed baryon
 fractions over all density contrasts and that the derived mean absolute values are significantly lower than IHE mean values: at $\delta=500$, the baryon fractions (gas fractions) are 11.5--13.4 \% (10.3--12 \%) and $\sim 20 \%$ (17 \%) respectively. We show that this is not due to the uncertainties on the outer slope $\beta$ of the gas density profile but is rather indicating that IHE masses are less reliable. Examining the shape of the baryon fraction profiles we find that cluster baryon fractions estimated from SLM follow a scaling law. Moreover, we do not find any strong evidence of increasing baryon (gas) fraction with temperature: hotter clusters do not have a higher baryon fraction than colder ones, neither do we find the slope $\beta$ to increase with temperature.

% The baryon distribution in clusters is expected to reflect the interplay %between galaxy formation and the intra-cluster medium. 
The absence of clear trends between $f_{b}$ and $M_{gas}/M_{*}$ with temperature is consistent with the similarity of baryon fraction profiles and suggests that non-gravitational processes such as galaxy feedback, necessary to explain the observed luminosity--temperature relationship, do not play a dominant  r\^ole in heating the intra-cluster gas on the virial scale.\\
\end{abstract}

\section{Introduction}

\hspace*{1em}
   Clusters of galaxies are fascinating objects because their observations
can in principle allow one to constrain the parameters of the standard
cosmological model. In particular, they are widely used as indicators of the
mean matter density of the universe. Galaxy clusters have been shown to 
harbour very large quantities of dark matter
since the pioneering work of Zwicky (1933), but its exact quantity, its spatial
distribution and above all its very nature are still awaiting answers. \\
\hspace*{1em}
   Clusters are the most massive objects
for which both the luminous baryonic mass (consisting of the X-ray emitting
intracluster gas and the visible part of galaxies) and the total gravitating 
mass can be estimated. Most often, the assumption of isothermal hydrostatic 
equilibrium (IHE) of the intra-cluster gas
within the dark matter potential well is adopted to derive the total mass of
clusters from X-ray observations, although many clusters exhibit obvious
substructures, both in the galaxy distribution and in the X-ray emission
morphology. \\
\hspace*{1em}
Beyond the classical $M/L$ ratio, clusters are at the center of new 
cosmological tests of the mean density, which are different in spirit
and which are more global. Partly because of this new perspective, 
general observational properties of clusters have been investigated in 
detail in recent years. These studies were triggered by analytical arguments as
well as numerical simulations which indicated that clusters might have similar
properties in their structure. A first means of determining the mean density
from clusters is to use their abundance as well as their 
relative evolution with redshift (Oukbir \& Blanchard 1992~; Bartlett 1997). A further important property of clusters is that their
baryon fraction $\fb$ is expected to be identical (White et al. 1993),
reflecting the universal baryonic content of the universe. As
primordial nucleosynthesis calculations provide very strong constraints on the
value of the baryonic density parameter $\omb$, determining the baryonic
fraction in galaxy clusters allows to derive the matter density
parameter $\omz = \omb/\fb$. This surmise, when applied to a set of clusters, leads to a high mean baryon
fraction $\fb$, of the order of $20 \h^{-3/2}$\% (David et al. 1995~, hereafter D95;
White \& Fabian 1995~; Cirimele et al. 1997~;
Evrard 1997). 
Consequently, the critical value
$\omz = 1$ is disfavored (as the primordial nucleosynthesis is indicative of
$\omb = 0.0776 \h^{-2}\:\eta_{10}/5.3 \pm 7\%$,
one obtains $\omz \sim 0.4$). White et al. (1993) have
reviewed this critical issue in the case of the Coma cluster. \\
\hspace*{1em}
   Some caution is necessary though, since there exists an appreciable 
dispersion in the range of published baryon fractions.
This scatter may be due to intrinsic dispersion in baryon fractions of different
clusters. If real, it is important to understand the origin of such a scatter.
However, Evrard (1997) 
did not find any convincing evidence for a significant
variation in the baryon fraction from cluster to cluster.
Such a result is in contrast with  Loewenstein \& Mushotzky (1996) and D95. These latter authors, 
from their study of ROSAT PSPC observations
of a sample of groups and clusters of galaxies, have found a correlation 
between the gas fraction and the gas temperature, breaking the simplest self-similar picture (the  different conclusion of Evrard could be due to the limited 
range of temperatures he used). A possible explanation for such variations, if real, could in principle be the development of a segregation
between baryons and dark matter occurring during the cluster collapse, operating
more efficiently in massive clusters. However, this mechanism
has been shown by White et al. (1993) to be insufficient to significantly
enhance the baryon fraction and it is therefore
unlikely that such a phenomenon could lead to a
substantial scatter in baryon fractions. Another possibility is that in poor
clusters and groups, a part of the gas has been swept away in the shallow
dark matter potential well by galactic winds, being thus less concentrated than
in massive clusters. This scenario would also be consistent with the claim 
that the gas to stellar mass ratio increases 
monotonically with the temperature of the cluster (David et al. 1990, hereafter D90). Finally, a further 
possibility is that mass estimates 
are not accurate and that a systematic bias exists with temperature.
In any case, D95 derived this correlation
from a very reduced set of objects (7 clusters and 4 groups) and it would
deserve further investigation based on a larger sample. \\
\hspace*{1em}
   As a consequence, it was one of our aims to address these questions 
with improved statistics. Moreover,
in the baryon problem, the reliability of mass estimates is
rather crucial and assumptions such as equilibrium and isothermality may
introduce systematic differences in the results that we wish to examine in detail. The validity of mass estimates has been questioned
by Balland \& Blanchard (1997). We have therefore taken the opportunity 
of this study to perform a comparison between the standard mass estimate
based on the IHE \bmodel and an alternative method derived from
scaling arguments and numerical simulations including gas physics (see section 3.3), hereafter called the scaling law model (SLM). \\
\hspace*{1em}
   In this paper, we present an analysis of a sample of 26 galaxy clusters and
7 groups taken from the literature. We required that optical data were available for our objects and searched for a precise information on the
galaxy spatial distribution and luminosity function, on the X-ray temperature
and on the gas density profile, in order to be able to build up the density and mass
profiles for galaxies, gas and dark matter. This allows to compute properly the baryon
fraction rather than only the gas fraction as is often done. This is especially important for low mass objects, in which the stellar component is generally believed to be relatively more important. Our sample comprises clusters with temperature from 1 to 14 keV, and therefore  allows us to investigate several interesting quantities beyond gas and baryon fraction, like the mass to light ratio and the ratio of galaxy baryonic mass to gas mass (possibly providing important constraints on galaxy formation),  over a wide range
of temperatures. All the data used here come from the literature, with
the exception of Abell 665, for which we have analysed an archival ROSAT
image to obtain the gas density profile. In fact, this cluster has already
been studied from Einstein data by two teams (Durret et al. 1994~; Hughes \&
Tanaka 1992), finding in each case a surprisingly very high gas fraction
(respectively $\simeq$ 50\% and 33\%). We will see this cluster provides a
striking example of the differences  in mass determinations. \\
\hspace*{1em}
  The sample is presented in section 2.  The methods to compute the various quantities for each cluster in  the sample is presented in section 3 and the results are presented in Section 4.  In section 5, we examine the trend with temperature for several quantities.
\\
\hspace*{1em}
   In all the present study, we assumed a $H_0 = 50 \km \s^{-1} \mpc^{-1}$ and
$q_0 = 0.5$ cosmology.

\section{The sample}

\begin{table*}[!h]
\caption[]{X-ray data.
	 Parameters refer to Eq.~9~; $\nez$ is the central electron density.
	  Temperature error bars of clusters are computed from different reliable
         references (taking into account or not a cooling flow)
         as the maximum of two estimates~: the dispersion among
         the measures and the quadratic mean of the quoted uncertainties (this procedure produces large uncertainty when there exists a possibility  for strong temperature gradient).
	 They are given at a 90\% confidence level, multiplying when necessary
         $1 \sigma$ errors by 1.64 (the errors of Fornax and RXJ1340.6 given by
         I96 and PA94, which confidence levels are not stated, are assumed to be
         $1 \sigma$). \\
	 $^{(*)}$ Fornax and HCG62 have got a two-component gas density profile,
	 the one in the second line corresponding to the nucleus.
	 When no limiting radius $\rxl$ is given, we assumed
	 $\rxl$ (Mpc) $\simeq \tx/3$ (keV), a relation calibrated on other clusters.}
\begin{flushleft}
\begin{tabular}{|cc|cc|ccccc|}
\noalign{\smallskip}
\hline
\noalign{\smallskip}
name & $z$ & $\tx$ & ref & $\rxl$ & $\beta$ & $\rcx$ & $\nez$ & ref  \\
~ & ~ & (keV) & ~ & (Mpc) & ~ & (Mpc) & ($10^{-3} \cm^{-3}$) & ~ \\
\noalign{\smallskip}
\hline
\noalign{\smallskip}
A76             & 0.0416  & $1.5 ^{+2.8}_{-0.9}$    & D93   & ~    & 0.58  & 0.356 & 1.07  & CNT97 \\
A85             & 0.0518  & $6.1 \pm 0.44$          & MF98  & 1.4  & 0.497 & 0.06  & 6.5   & P97 \\
A119            & 0.0440  & $5.62 \pm 0.61$         & W00   & ~    & 0.56  & 0.378 & 1.18  & CNT97 \\
A401            & 0.0748  & $8.3 \pm 1.16$          & MF98  & 2.   & 0.606 & 0.285 & 4.31  & BC96, EVB96 \\
A426 (Perseus)  & 0.0183  & $6.79 \pm 0.76$	    & F98   & 2.5  & 0.727 & 0.416 & 2.89  & CK97 \\
A576            & 0.0381  & $4.02 \pm 0.31$         & W00   & 1.5  & 0.64  & 0.24  & 2.47  & MG96 \\
A665            & 0.1816  & $8.26 \pm 1.29$         & D93   & 2.36 & 0.763 & 0.44  & 2.85  & this study\\
A1060 (Hydra I) & 0.0114  & $3.1 \pm 0.14$          & T96   & ~    & 0.61  & 0.094 & 5.00  & LM96 \\
A1377           & 0.0514  & $2.6 ^{+0.7}_{-1.2}$    & WJF97 & ~    & 0.46  & 0.188 & 0.75  & CNT97 \\
A1413           & 0.1427  & $7.32 \pm 1.10$	    & W00   & ~    & 0.62  & 0.156 & 8.00  & CNT97 \\
A1656 (Coma)    & 0.0232  & $8.11 \pm 0.79$	    & D93   & 4.   & 0.75  & 0.42  & 2.89  & BHB92 \\
A1689           & 0.1810  & $9.23 \pm 1.05$         & W00   & ~    & 0.72  & 0.205 & 14.0  & CNT97 \\
A1775           & 0.0696  & $3.69 \pm 3.03$         & W00   & ~    & 0.58  & 0.174 & 2.73  & CNT97 \\
A2029           & 0.0767  & $8.7 \pm 0.75$          & MF98  & 2.3  & 0.682 & 0.251 & 6.5   & BC96, EVB96 \\
A2052           & 0.0348  & $3.03 \pm 0.23$	    & W00   & ~    & 0.67  & 0.142 & 4.43  & CNT97 \\
A2063           & 0.0337  & $3.52 \pm 0.64$         & W00   & ~    & 0.50  & 0.074 & 5.9   & CNT97 \\
A2163           & 0.201   & $14.6 \pm 1.21$         & EAB95 & 4.6  & 0.62  & 0.305 & 6.65  & EAB95 \\
A2199           & 0.0303  & $4.27 \pm 0.28$         & W00   & ~    & 0.62  & 0.117 & 6.5   & CNT97 \\
A2218           & 0.175   & $6.84 \pm 2.83$	    & W00   & 2.1  & 0.71  & 0.276 & 5.54  & SK96 \\
A2256           & 0.058   & $6.98 \pm 0.44$	    & W00   & 3.   & 0.795 & 0.54  & 2.51  & HBN93 \\
A2634           & 0.0312  & $3.27 \pm 0.28$	    & W00   & ~    & 0.58  & 0.32  & 0.89  & CNT97 \\
A2657           & 0.0414  & $3.81 \pm 0.40$         & W00   & ~    & 0.52  & 0.124 & 4.4   & CNT97 \\
A2670           & 0.0745  & $3.73 \pm 0.70$         & W00   & ~    & 0.70  & 0.174 & 3.83  & CNT97 \\
AWM7            & 0.0176  & $3.79 \pm 0.36$         & W00   & 1.25 & 0.53  & 0.102 & 6.47  & NB95 \\
Hydra A (A780)  & 0.0522  & $3.56 \pm 0.37$         & W00   & 1.   & 0.7   & 0.145 & 6.5   & D90 \\
Fornax$^{(*)}$  & 0.0046  & $1.09 \pm 0.18$	    & I96   & 0.4  & 0.60  & 0.175 & 0.70  & I96 \\
   ~            & ~	  & ~			    & ~     & ~    & 0.51  & 0.007 & 19.6  & ~ \\
HCG62$^{(*)}$   & 0.0138  & $0.96 \pm 0.07$	    & PB93  & 0.36 & 0.38  & 0.06  & 1.53  & PB93 \\
   ~            & ~	  & ~			    & ~     & ~    & 0.9   & 0.019 & 15.6  & ~ \\
HCG94           & 0.04218 & $2.75 ^{+0.9}_{-0.4}$   & EMW95 & 1.   & 0.58  & 0.17  & 3.0   & EMW95 \\
NGC533          & 0.0172  & $1.05 ^{+0.05}_{-0.09}$ & MD96  & 0.58 & 0.69  & 0.237 & 0.7   & MD96 \\
NGC2300         & 0.0076  & $0.97 ^{+0.11}_{-0.08}$ & DM96  & 0.33 & 0.41  & 0.057 & 3.11  & DM96 \\
NGC4261         & 0.0088  & $0.85 ^{+0.21}_{-0.16}$ & DM95  & 0.6  & 0.31  & 0.026 & 2.9   & DM95 \\
NGC5044         & 0.0087  & $0.98 \pm 0.04$	    & D94   & 0.4  & 0.53  & 0.028 & 7.9   & D94 \\
RXJ 1340.6+4018 & 0.171   & $0.92 \pm 0.08$	    & PA94  & 0.4  & 1.    & 0.181 & 2.5   & PA94 \\
\noalign{\smallskip}
\hline
\end{tabular}
\end{flushleft}
\end{table*}

\begin{table*}[!h]
\caption[]{Optical data.
	 The first part corresponds to the galaxy spatial distribution
	 and the second part to the luminosity function (Eq.~5). \\
	 $^{(*)}$ TOP~: type of optical profile~: (1) stands for a King form (Eq.~2),
	 (2) for a de Vaucouleurs form (Eq.~3), (3) for an integrated
	 luminosity profile (Eq.~7) and (4) for a de Vaucouleurs projected
	 luminosity density profile (cf Eq.~3). \\
	 $^{(*)}$ $\sigma_0$~: (1) and (2)~: galaxies$ \mpc^{-2}$~;
	 (3)~: $L_0$ ($10^{11} \lbsun$)~;
	 (4)~: $\sigma_{\rm L0}$ ($10^{11} \lbsun \mpc^{-2}$). \\
	 $^{(*)}$ $\rc$~: for (2) and (4), corresponds to $\rv$. \\
	 $^{(*)}$ $\epsilon$~: for (2) and (4), corresponds to $\gamma$.}
\begin{flushleft}
\begin{tabular}{|c|cccccc|ccc|}
\noalign{\smallskip}
\hline
\noalign{\smallskip}
name & TOP$^{(*)}$ & $\sigma_0^{(*)}$ & $\rc^{(*)}$ & $\epsilon^{(*)}$ & $\mblim$ & ref & $\mbst$ & $\alpha$ & ref \\
~ & ~ & ~ & (Mpc) & ~ & ~ & ~ & ~ & ~ & ~ \\
\noalign{\smallskip}
\hline
\noalign{\smallskip}
A76     & (2) & 39.8   & 0.485   & 0.678 & -18.77 & CNT97	 & -21.00 & 1.25 & CNT97 \\
A85     & (1) & 253    & 0.518	 & 1	 & -17.76 & M84 	 & -21.17 & 1.13 & OH89 \\
A119    & (2) & 214    & 0.023   & 0.271 & -19.77 & CNT97	 & -20.98 & 1.25 & CNT97 \\
A401    & (1) & 165    & 0.4	 & 1	 & -18.27 & D78b	 & -20.96 & 1.25 & D78a \\
A426    & (1) & 209    & 0.308	 & 1	 & -17.34 & KS83	 & -20.93 & 1.25 & ~ \\
A576    & (1) & 250    & 0.53	 & 1	 & -16.26 & MG96	 & -20.58 & 1.18 & MG96 \\
A665    & (1) & 168    & 0.5	 & 1	 & -18.27 & D78b	 & -20.21 & 1.25 & D78a \\
A1060   & (1) & 228    & 0.180   & 1	 & -16.62 & BO78	 & -20.93 & 1.25 & ~ \\
A1377   & (2) & 233    & 0.170   & 0.751 & -19.47 & CNT97	 & -21.00 & 1.25 & CNT97 \\
A1413   & (2) & 281    & 0.05	 & 0.5   & -19.97 & CNT97	 & -20.59 & 1.25 & CNT97 \\
A1656   & (1) & 200    & 0.34	 & 1	 & -18.25 & KG82	 & -20.93 & 1.25 & ~ \\
A1689   & (2) & 140    & 0.059   & 0.43  & -20.77 & CNT97	 & -21.66 & 1.25 & CNT97 \\
A1775   & (2) & 205    & 0.007   & 0.227 & -19.57 & CNT97	 & -21.44 & 1.25 & CNT97 \\
A2029   & (1) & 198    & 0.35	 & 1	 & -18.27 & D78b	 & -21.14 & 1.17 & OH89 \\
A2052   & (2) & 175    & 0.105   & 0.553 & -19.57 & CNT97	 & -20.35 & 1.25 & CNT97 \\
A2063   & (2) & 153    & 0.19	 & 0.767 & -19.47 & CNT97	 & -20.30 & 1.25 & CNT97 \\
A2163   & (3) & 139    & 0.532	 & ~	 & ~	  & SN97	 & ~	  & ~	 & ~ \\
A2199   & (2) & 442    & 0.008   & 0.264 & -19.27 & CNT97	 & -20.17 & 1.25 & CNT97 \\
A2218   & (1) & 387    & 0.4	 & 1	 & -18.27 & D78b	 & -20.52 & 1.25 & D78a \\
A2256   & (1) & 138    & 0.49	 & 1	 & -18.27 & D78b	 & -21.27 & 1.12 & OHJ87 \\
A2634   & (2) & 508    & 0.008   & 0.252 & -19.27 & CNT97	 & -19.75 & 1.25 & CNT97 \\
A2657   & (2) & 648    & 0.01	 & 0.345 & -19.57 & CNT97	 & -20.86 & 1.25 & CNT97 \\
A2670   & (2) & 245    & 0.046   & 0.393 & -19.47 & CNT97	 & -21.00 & 1.25 & CNT97 \\
AWM7    & (3) & 6.25   & 0.102	 & ~	 & ~	  & B84 	 & ~	  & ~	 & ~ \\
Hydra A & (3) & 6.07   & 0.146   & ~	 & ~	  & D90 	 & ~	  & ~	 & ~ \\
Fornax  & (1) & 250    & 0.337   & 1	 & -13.20 & F89 	 & -20.82 & 1.32 & FS88 \\
HCG62   & (3) & 3.07   & 0.180   & ~	 & ~	  & ZM98	 & ~	  & ~	 & ~ \\
HCG94   & (3) & 7.72   & 0.17    & ~	 & ~	  & HKA89, EMW95 & ~	  & ~	 & ~ \\
NGC533  & (3) & 2.64   & 0.140   & ~	 & ~	  & ZM98	 & ~	  & ~	 & ~ \\
NGC2300 & (3) & 3.26   & 0.245   & ~	 & ~	  & G93          & ~	  & ~	 & ~ \\
NGC4261 & (3) & 5.82   & 0.163	 & ~	 & ~	  & N93 	 & ~	  & ~	 & ~ \\
NGC5044 & (1) & 331    & 0.188	 & 1	 & -15.34 & FS90	 & -20.93 & 1.25 & ~ \\
RXJ     & (4) & 3449   & 0.00524 & 0.665 & ~	  & PA94	 & ~	  & ~	 & ~ \\
\noalign{\smallskip}
\hline
\end{tabular}
\end{flushleft}
\end{table*}

\hspace*{1em}
   We looked in the literature for objects studied thoroughly enough to
allow us to compute the baryonic mass in galaxies, the mass in gas and in dark matter at any
radius. This  is quite not a refinement, since both the baryon fraction
and the galaxy to gas  mass ratio can vary very rapidly with radius, as
will be seen in the next section.
We therefore needed detailed information,
which drastically reduced the possible number
of objects that could be included in the sample. When a same object was studied by
several teams, we applied  straightforward selection criteria~: for spatial X-ray
data, for instance, we systematically prefer ROSAT observations, because of
its improved spatial resolution and sensitivity, whereas for X-ray temperatures,
Ginga and ASCA satellites were preferred to Einstein MPC,  most of which come from the catalogue of David
et al. (1993).  Recently, it has been noticed that cluster luminosities 
and temperatures might change noticeably when the central cooling flow emission is
removed (Markevitch 1998~; Arnaud \& Evrard 1999). It is not clear which temperatures are to be used (especially when using a mass-temperature
relationship derived from numerical simulations). In order to keep our sample
as homogenous as possible, we did not use cooling flow-corrected temperatures
which are not always available. Furthermore Markevitch (1998) found that
temperatures corrected for central emission are in the mean 3\% larger,
which will be of weak consequence in our average quantities. 
However,
 our treatment of the uncertainties on temperatures 
 leads to large error bars when a large dispersion in measured temperatures exists (see Table 1), as for 
instance  in the presence of strong cooling flow. \\
\hspace*{1em}
   In some cases, optical
data may be very uncertain because of projection effects and magnitude
limitations, especially for groups whose galaxy membership is sometimes tricky
to establish. However, we tried to identify objects for which data are 
reasonably
reliable and we derived mean dynamical quantities for this sub-sample as well.
Finally, it must be emphasised that the X-ray limiting radius at which baryon
fractions are estimated is a crucial parameter, since both the galactic mass
derived from a King profile and the X-ray gas mass given by the Hubble-King model
diverge respectively for $\epsilon \leq 1$ and $\beta \leq 1$ (the definition is given in §3), requiring that 
they
be truncated. It is also
important that the baryon fractions of different clusters be computed at
an equivalent scale in order to test the scaling hypothesis and 
if statistical conclusions are to be brought out from them, \textit{i.e.}
that we use the radius containing the same overdensity, while information is 
actually available only up to the X-ray limiting radius $\rxl$ which 
primarily depends on the
characteristics of the observations (detector sensitivity, 
integration time...). \\
\hspace*{1em}
   X-ray and optical data are summarised in Tables 1 and 2, using a Hubble 
constant $\h = 1$. Notes on clusters which required a special treatment due to an incompleteness of data can be found at the end. Optical luminosities are given in the blue band.
 When the blue luminosity was not available, we used the following 
colors, corresponding to standard values for elliptical
galaxies~: B-V = 0.97, V-F = 0.76, r-F = 0.58 (Schneider et al. 1983)
and R = F (Lugger 1989).\\

\subsection{The case of Abell 665}

\begin{figure}[!ht]
\includegraphics[scale=0.35,angle=-90]{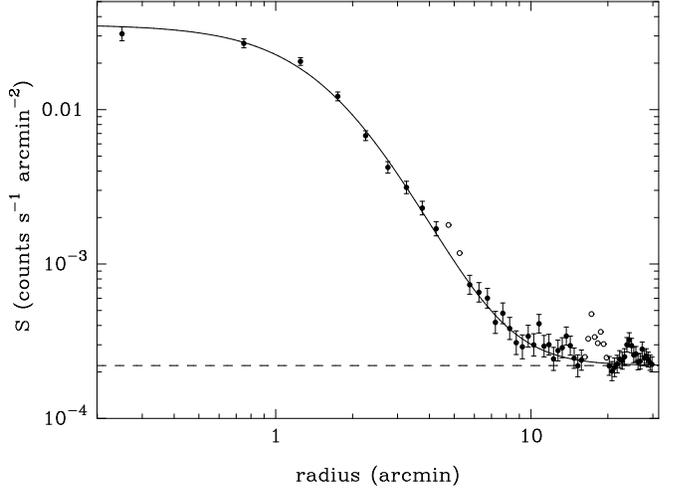}
\caption[]{Surface brightness profile of the intracluster gas in A665, in the
           ROSAT bands R4 to R7 (0.44 to $2 \kev$). 
           Points represented by an empty circle have been excluded from the fit
           because at these radii some background or foreground X-ray sources
           appear in the map. The dashed line is the fitted background level.
\label{fig:a665}}
\end{figure}

\hspace*{1em}
   This cluster is one for which large baryon fraction estimates 
have been published in the literature. As these are surprisingly high, we have found interesting to 
re-analyse this cluster 
using a ROSAT archival image and the calibration routines of Snowden
et al. (1994). We found that the gas surface brightness
profile is well fitted by a Hubble-King law, and the X-ray emission can
be traced out to a very large radius. The background level, which has been
fitted together with the other parameters, is estimated with comfortable 
confidence.
Spherical symetry was assumed to derive the surface brightness profile
in $0.5 \arcm$ wide annuli, although the X-ray map shows significant departure
from sphericity~; however, the effect of ellipticity on derived masses is known
to be negligible (Buote \& Canizares 1996).
The central electron volume density $\nez$ was computed by matching the
theoretical count rate with the $0.547 \counts \s^{-1}$ collected within a
$30 \arcm$
radius (after subtraction of the background), which amounts to solving:
\begin{eqnarray}
   \nez^2\:\alpha_{\rm E}\:D\:T^{-\frac{1}{2}} & & \int_{E_{\rm min}}^{E_{\rm max}}
      \frac{g(T,E)\:e^{-\frac{E}{kT}}\:e^{-\sigma(E) N_{\rm H}}\:A(E)}{E}\:dE \nonumber \\
   & & \int_0^{\thetamax} \paro 1 + \paro \frac{\theta}{\thetac} \parf ^2
      \parf ^{-3 \beta} \theta^2 d\theta \nonumber \\
   = 2 \pi\:S_0 & & \int_0^{\thetamax} \paro 1 + \paro \frac{\theta}{\thetac} \parf ^2
      \parf ^{(-3 \beta + \frac{1}{2})} \theta d\theta,
\end{eqnarray}
with $\alpha_{\rm E} = 1.02\:10^{-17}$ SI and the angular distance $D \simeq 812 \mpc$,
and where $A(E)$ stands for the energy dependence of the transmission 
efficiency.
The results of this analysis are the following (for the bands R4 to R7 of
ROSAT)~: $\beta = 0.763 \pm 0.023$, $\thetac = (112 \pm 5) \arcs$ (which
corresponds to $0.44 \mpc$ at the distance of A665),
$\nez = (2.85 \pm 0.38)\:10^{-3}~{\rm electrons}~\cm^{-3}$ and
$\rxl = 10~\arcm~(= 2.36 \mpc$) with a central surface brightness
$S_0 = (3.53 \pm 0.26)\:10^{-2} \counts \s^{-1} \arcm^{-2}$ and a background
surface brightness $b = (2.2 \pm 0.6)\:10^{-4} \counts \s^{-1} \arcm^{-2}$.
We used the gas temperature and foreground absorbing hydrogen column density
given by Hughes \& Tanaka (1992) from their Ginga analysis, together with the
formula of Mewe et al. (1986) for the Gaunt factor and that of Morrison
\& McCammon (1983) for the interstellar absorption cross section.
At $\rxl$, the inferred gas mass is $\mgas = (2.46 \pm 0.76)\:10^{14} \msun$ (at, which
is similar to the values found by Durret et al. (1994) and Hughes \& 
Tanaka (1992). The hydrostatic mass is
$M_{\rm hydro} = (1.60 \pm 0.24)\:10^{15} \msun$.
Our mass estimate from NFW's dark matter profile, computed with the EMN normalisation (see section below) is 
$M_{\rm dark} = (1.41 \pm 0.25)\:10^{15} \msun$,
and the resulting total mass is $M_{\rm SLM} = (1.68 \pm 0.33)\:10^{15} \msun$.
The baryon fraction amounts to respectively ($16.3 \pm 7.5$)\% and
($15.6 \pm 6.4$)\%. Hence A665 is a quite ordinary rich cluster whose baryon fraction
seems reasonable if compared to previous values. We have also compared our gas mass estimates for the whole sample with other  published analyses and found good agreement while the main differences are on $f_b$, coming from the estimation of total masses as will be discussed in §5.

\section{Analysis methods}
\begin{figure*}[!ht]
\begin{minipage}[t]{7.8cm}
\includegraphics[scale=0.293,angle=-90]{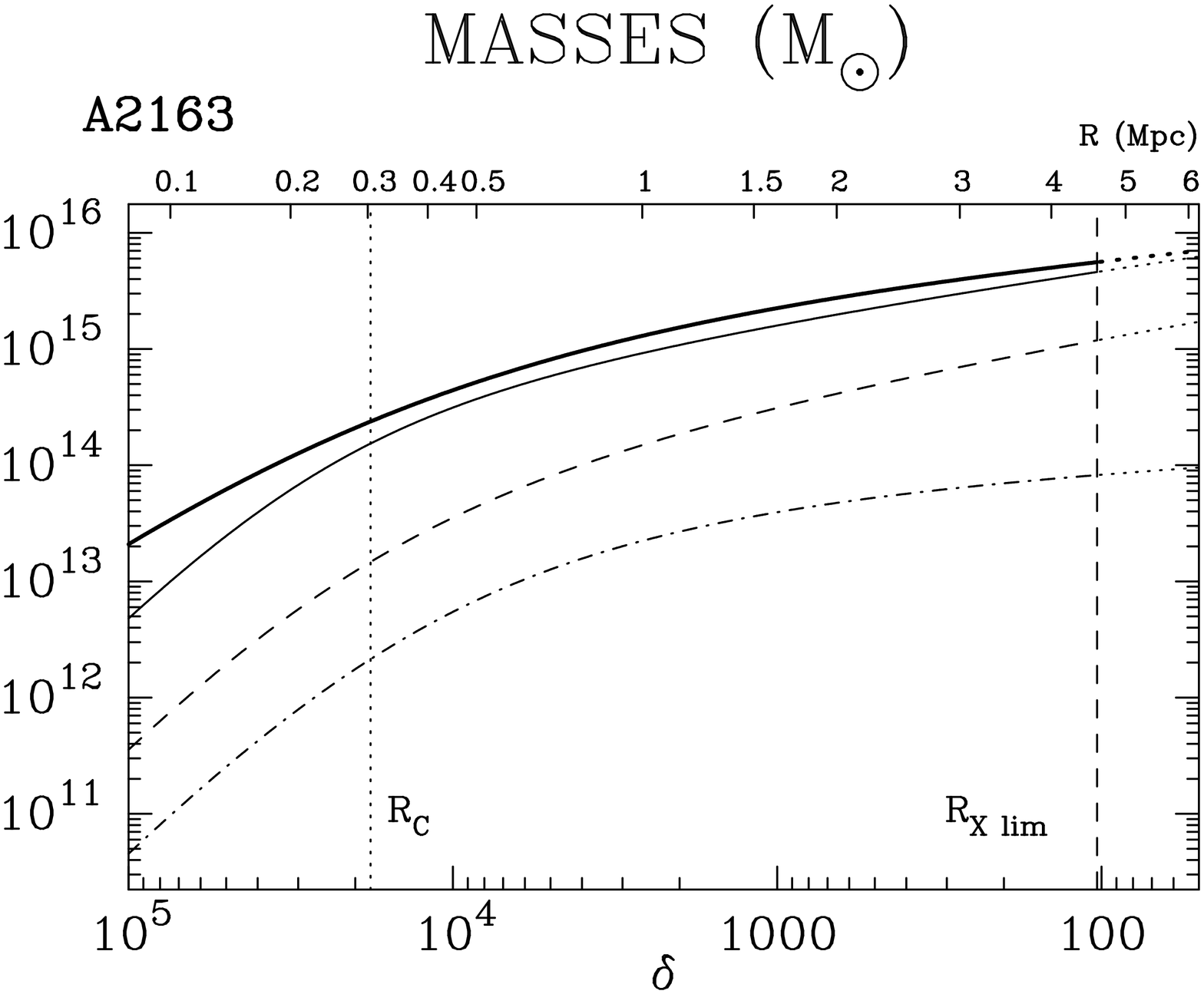}
\end{minipage}
\begin{minipage}[t]{7.8cm}
\includegraphics[scale=0.293,angle=-90]{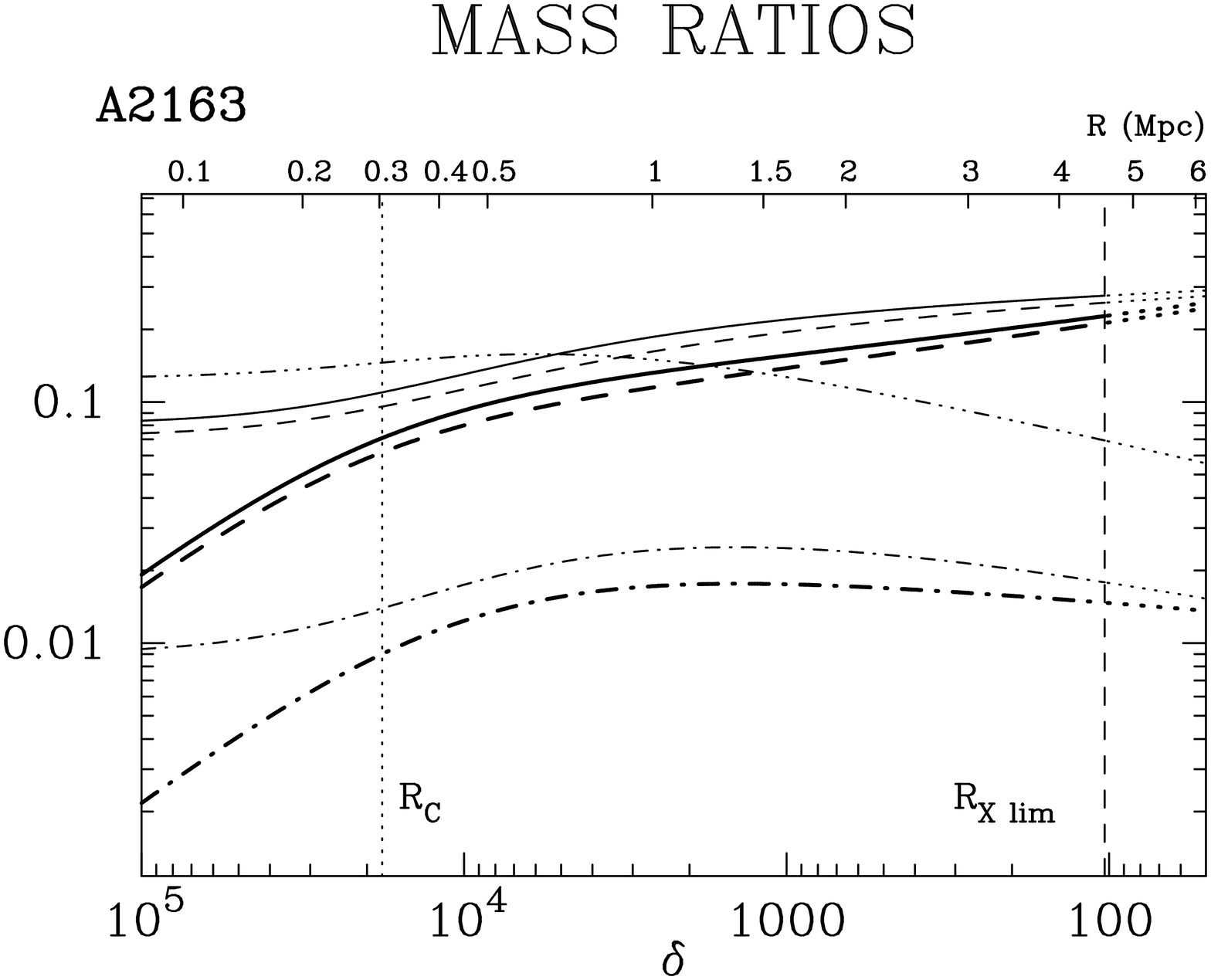}
\end{minipage}
\vspace*{-0.7cm} \\
\begin{minipage}[t]{7.8cm}
\includegraphics[scale=0.293,angle=-90]{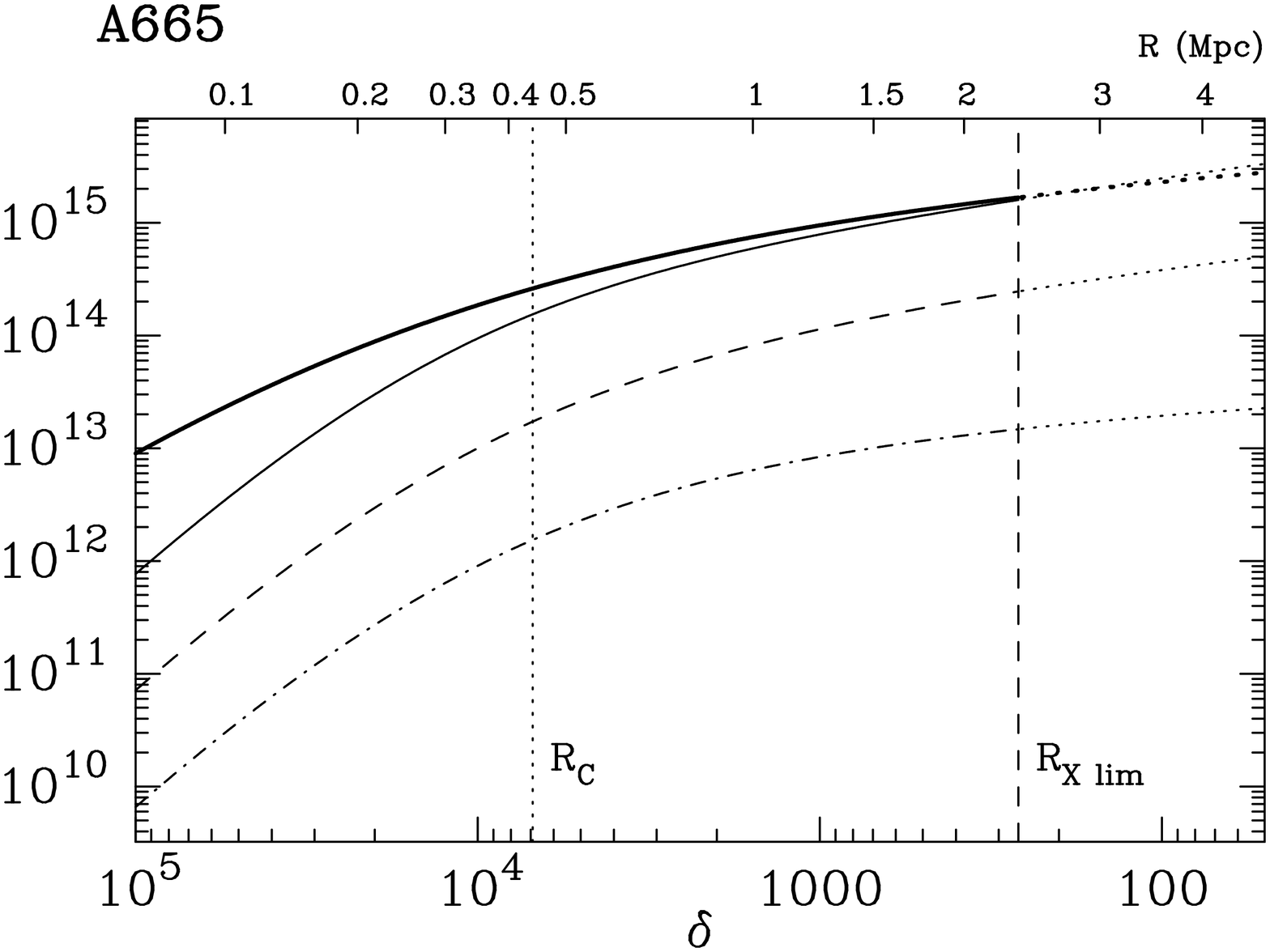}
\end{minipage}
\begin{minipage}[t]{7.8cm}
\includegraphics[scale=0.293,angle=-90]{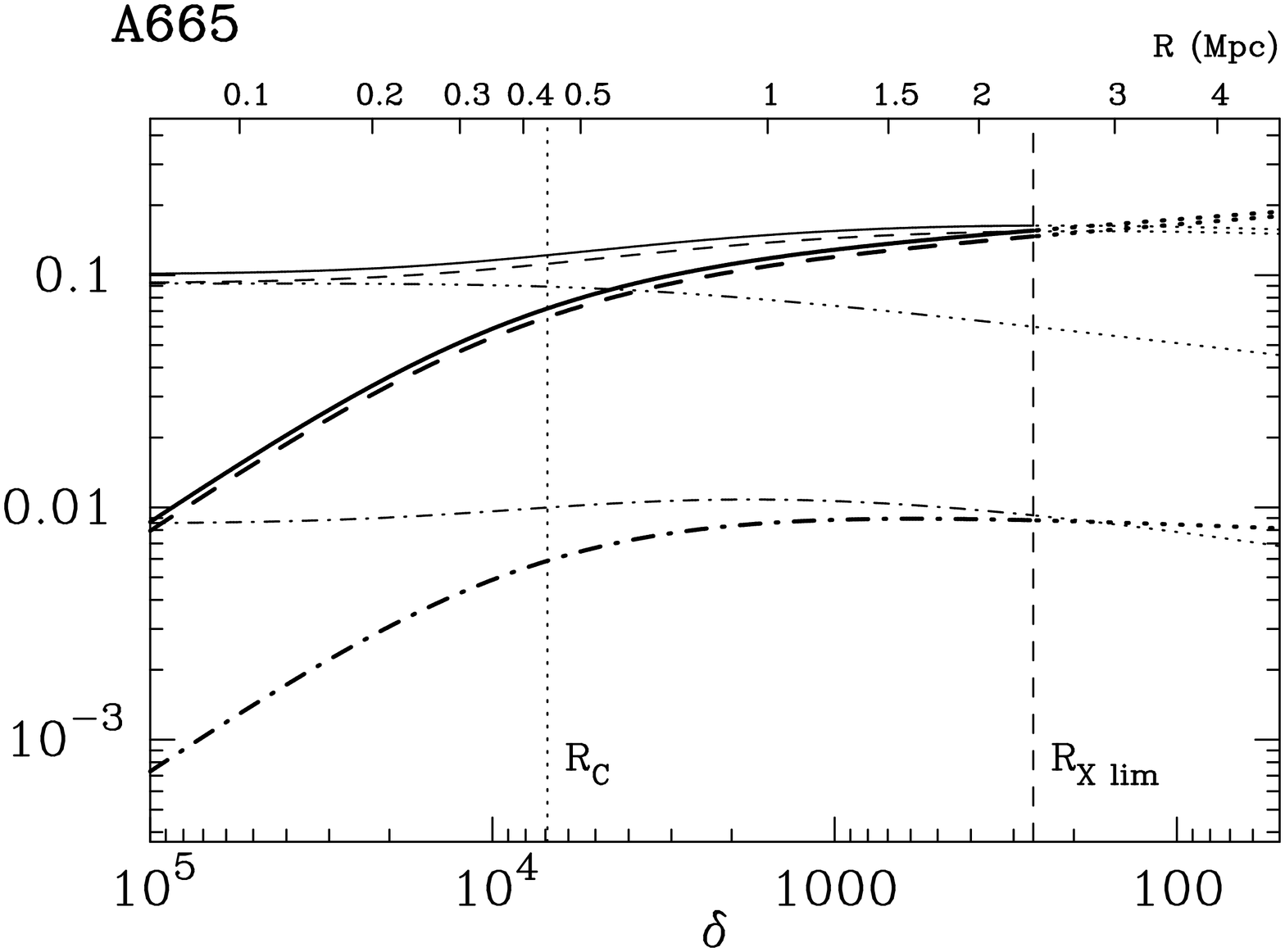}
\end{minipage}
\vspace*{-0.7cm} \\
\begin{minipage}[t]{7.8cm}
\includegraphics[scale=0.293,angle=-90]{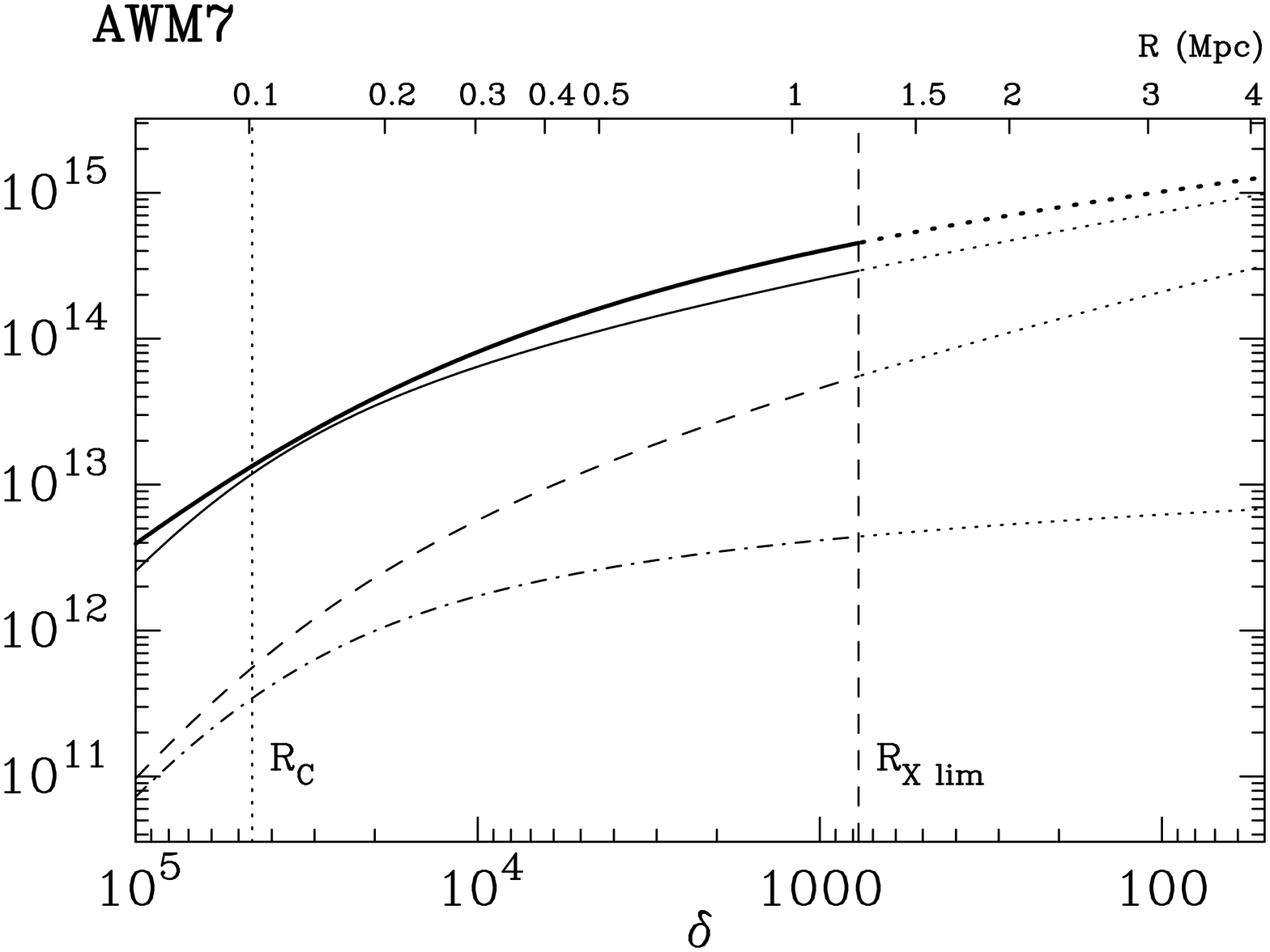}
\end{minipage}
\begin{minipage}[t]{7.8cm}
\includegraphics[scale=0.293,angle=-90]{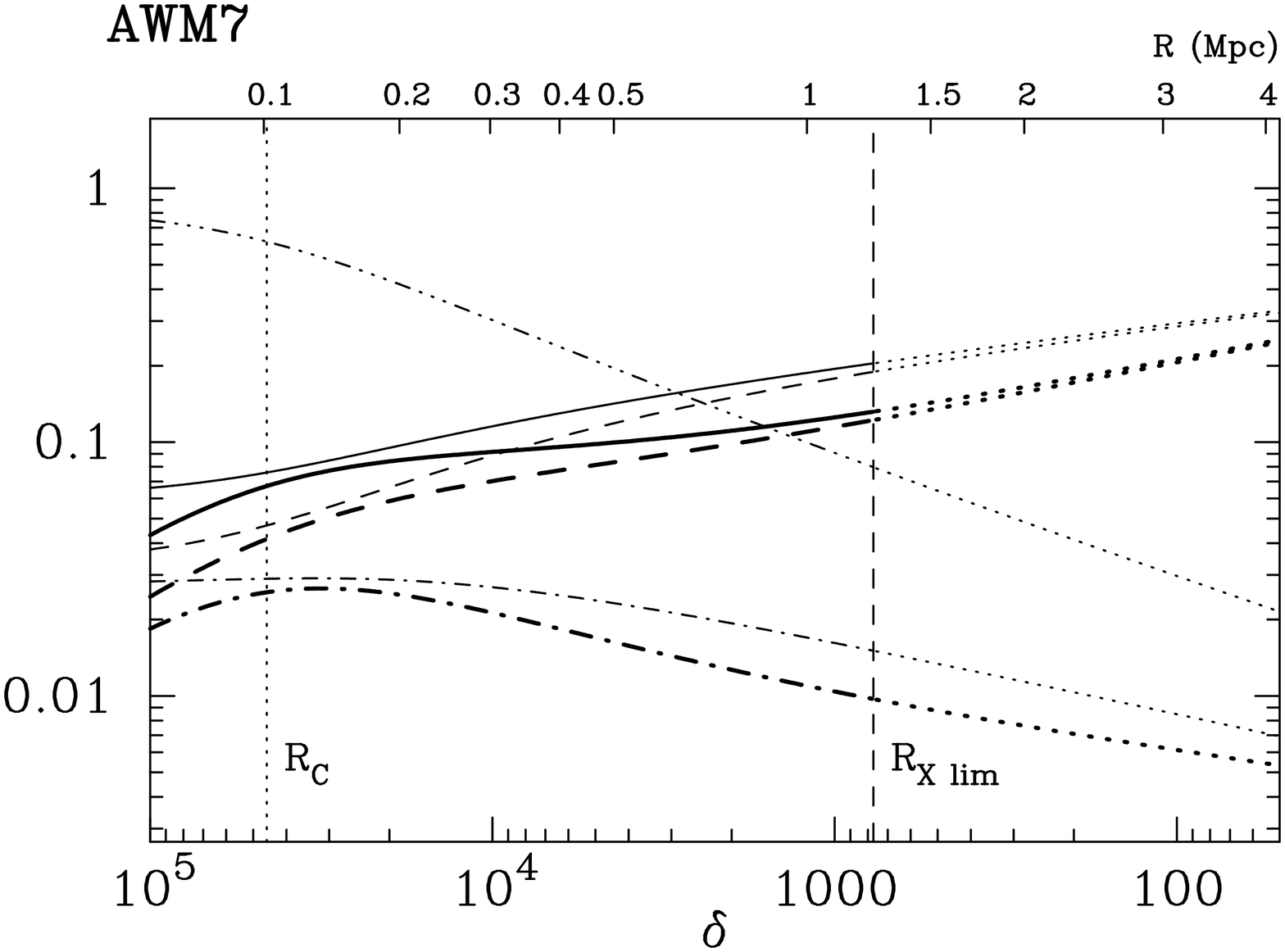}
\end{minipage}
\vspace*{-0.7cm} \\
\begin{minipage}[t]{7.8cm}
\includegraphics[scale=0.293,angle=-90]{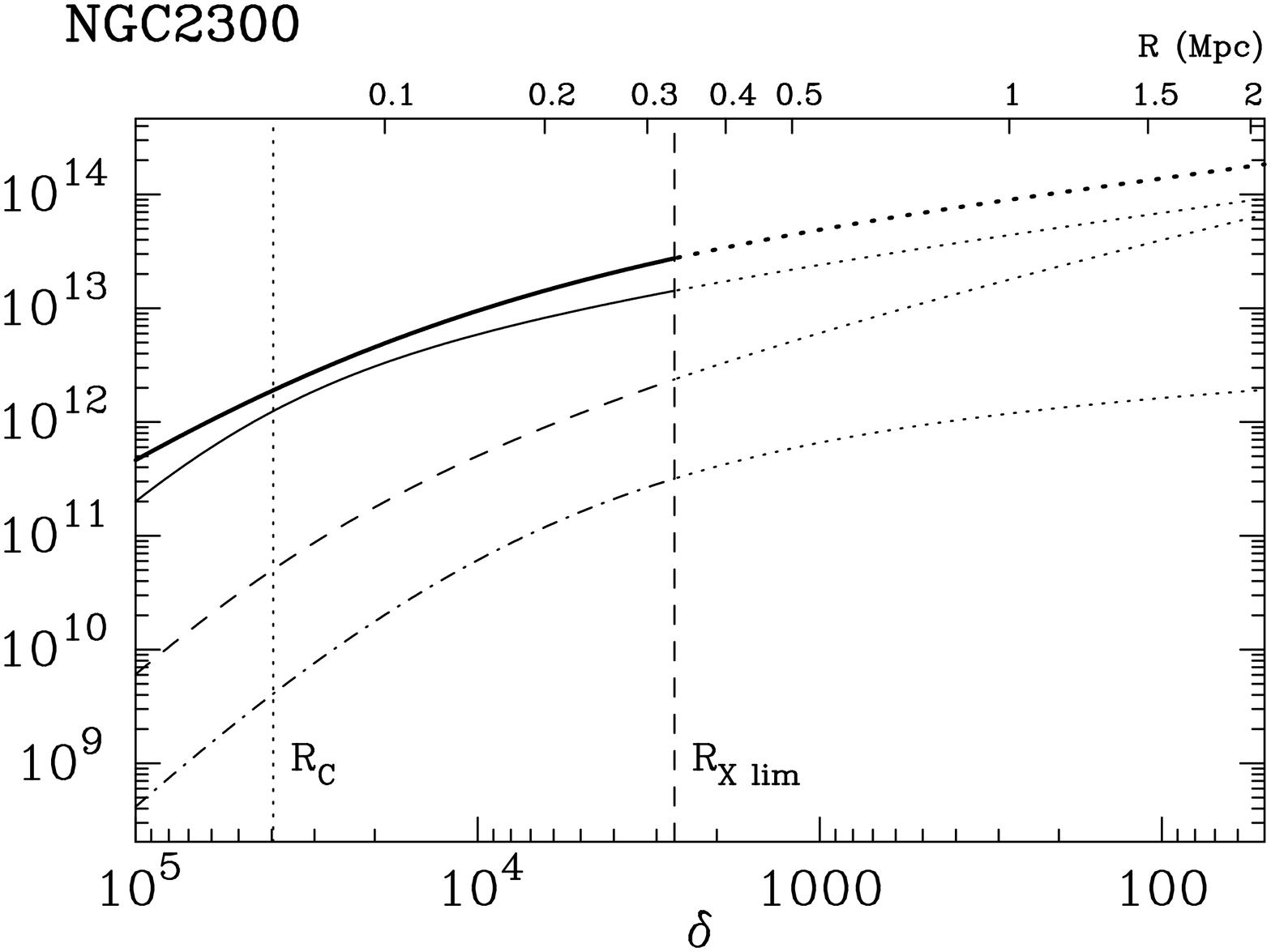}
\end{minipage}
\begin{minipage}[t]{7.8cm}
\includegraphics[scale=0.293,angle=-90]{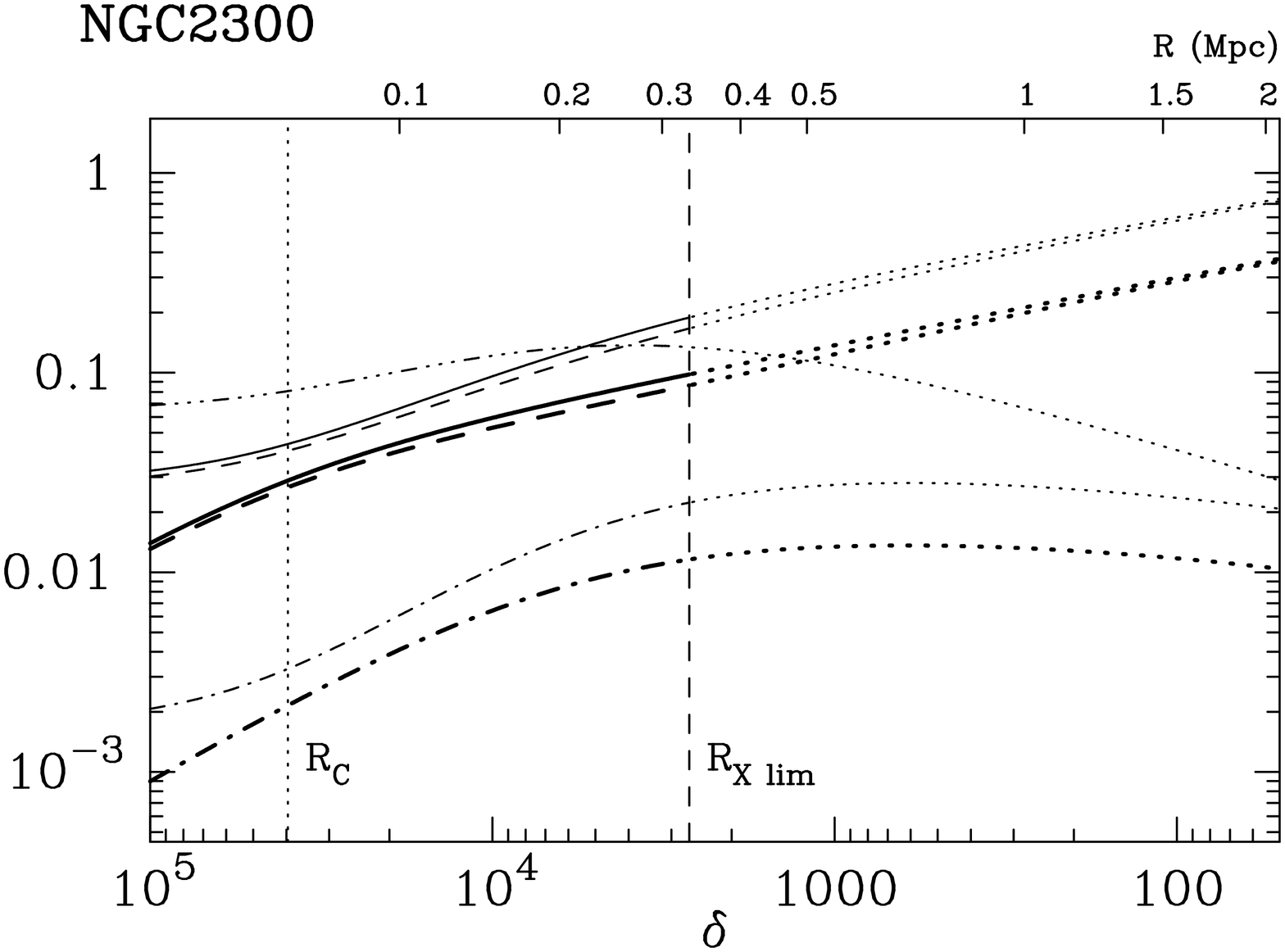}
\end{minipage}
\caption[]{Mass and mass ratio profiles for a few objects.
           The meaning of line styles is as follows~: 
           Left panels~: thick line~: SLM mass~;
           thin line~: IHE mass~;
           dashes~: gas mass~; dot-dashed line~: stellar mass. 
           Right panels~: thick lines~: baryon fraction (continuous),
           gas fraction (dashed) and stellar to total mass ratio
           (dot-dashed) in the SLM case (with EMN calibration)~; thin lines~: same quantities for the IHE
           model~; three-dots-dash~: stellar to gas mass ratio.\label{fig:profiles}}
%\vspace{-11ex}
\end{figure*}

% figures all_fb... et all_fgaz... ou all_logfb... et all_logfgaz...
\begin{figure*}[!ht]
\begin{minipage}[t]{8cm}
\includegraphics[scale=0.3,angle=-90]{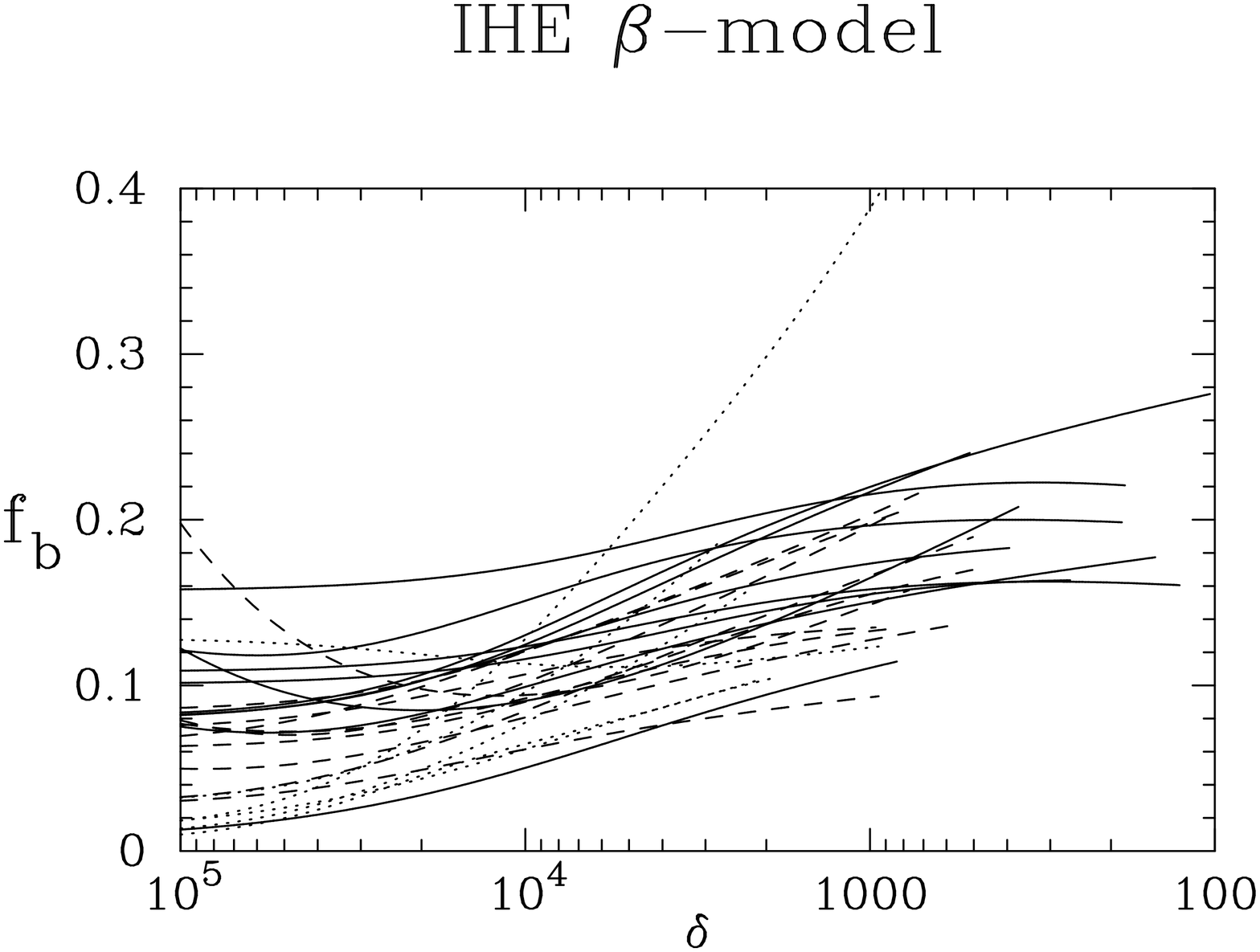}
\end{minipage}
\begin{minipage}[t]{8cm}
\includegraphics[scale=0.3,angle=-90]{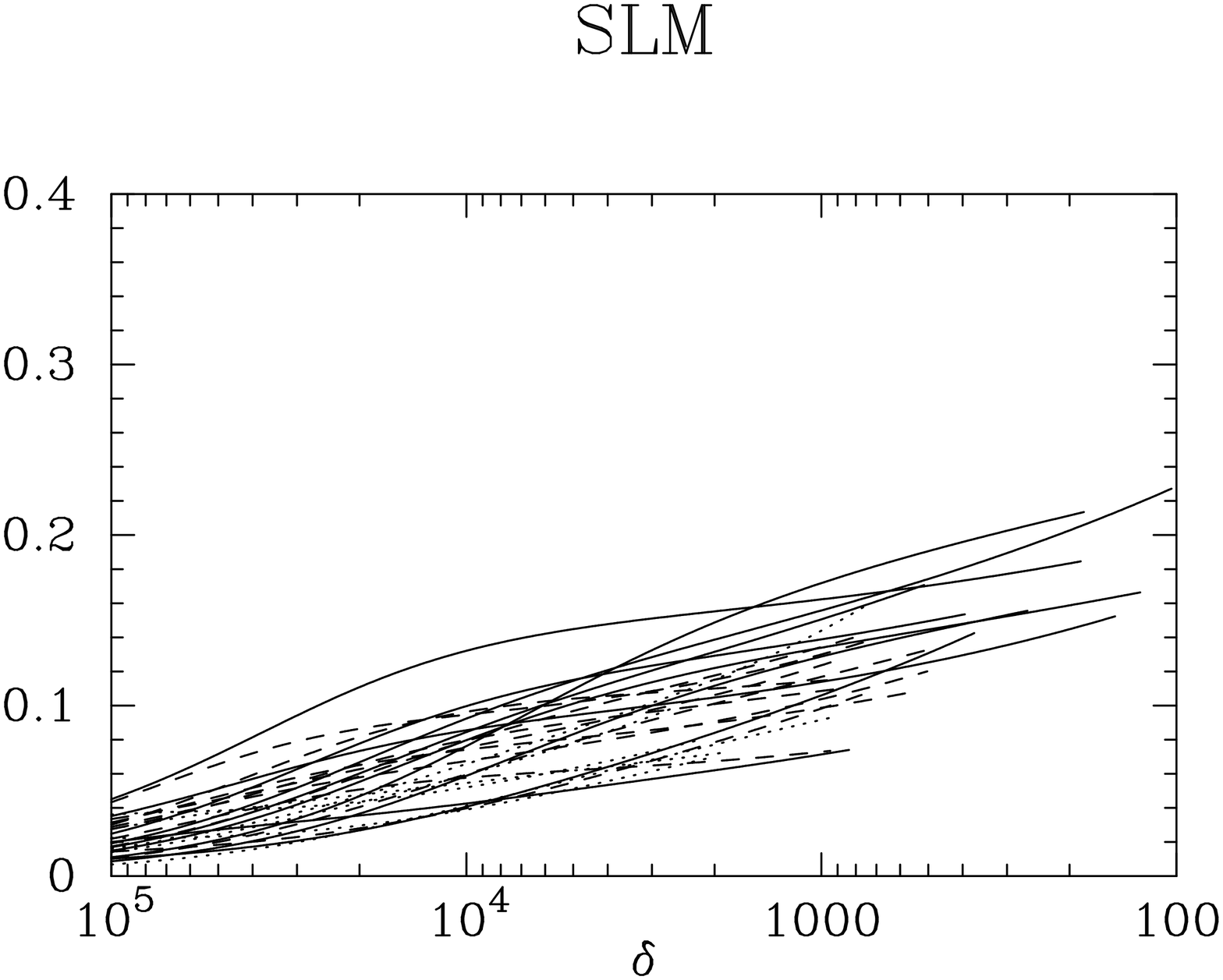}
\end{minipage}
\vspace*{-0.5cm} \\
\begin{minipage}[t]{8cm}
\includegraphics[scale=0.3,angle=-90]{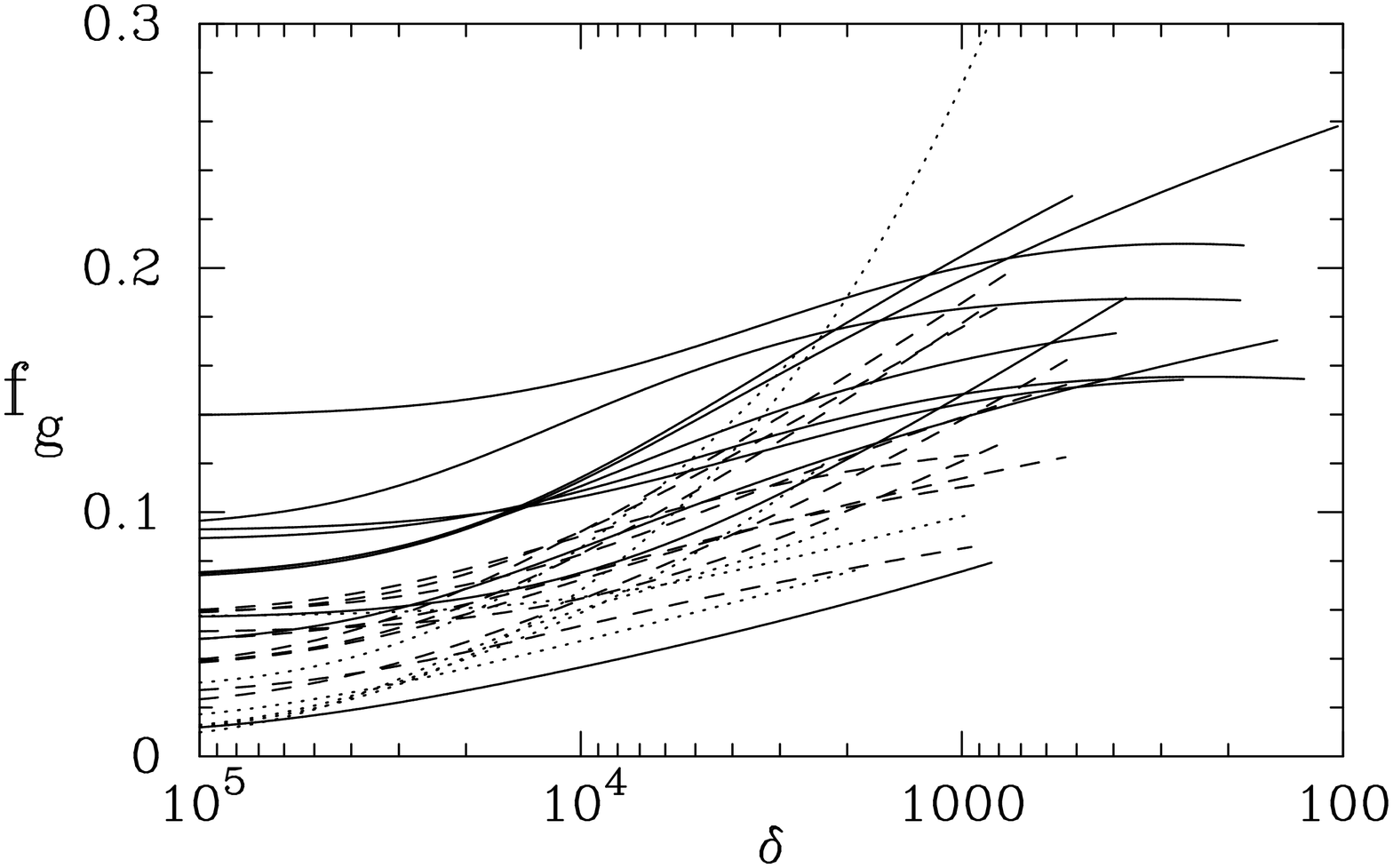}
\end{minipage}
\begin{minipage}[t]{8cm}
\includegraphics[scale=0.3,angle=-90]{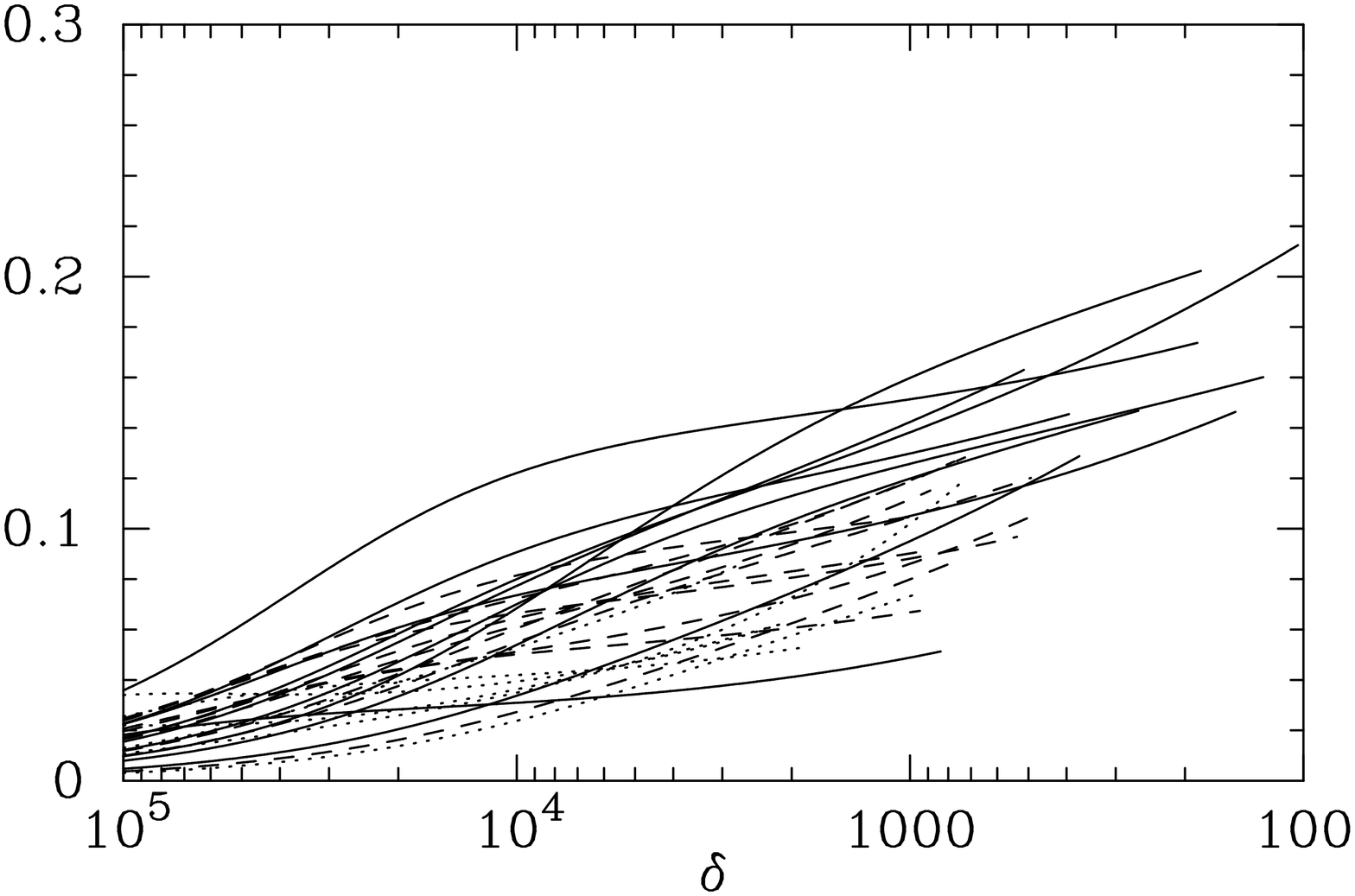}
\end{minipage}
\caption[]{Profiles of the baryon fraction and gas fraction
           as a function of mean overdensity for objects with
           the most reliable data. 
           Left panels show these profiles in the case of the hydrostatic
           assumption and right panels for mass estimates derived from
           NFW's dark matter profile, with EMN normalisation. 
           Groups ($\tx \leq 2 \kev$) are represented with dotted lines,
           cool clusters ($\tx \leq 5 \kev$) with dashed lines and
           hot clusters with continuous lines. The group with a very steeply
           rising baryon fraction in the IHE case has $\beta = 0.31$. \label{fig:tprofiles}}
\end{figure*}

\subsection{The stellar mass profile}

\hspace*{1em}
   The stellar matter content can be computed at any radius from the
cluster center using the projected number density profile of galaxies, their 
luminosity
function and a mass to light ratio for the stellar population calibrated on the
observation of nearby galaxies. Most often, the density profile is fitted by the
common King form~:
\begin{equation}
\label{eq:king}
   \sigma_{\rm gal}(r) = \sigma_0 \paro 1 + \paro \frac{r}{\rc} \parf ^2 \parf
   ^{-\epsilon}~{\rm with}~\epsilon = 1,
\end{equation}
where $\rc$ is the galactic core radius. The case $\epsilon = 1$ is an
approximation to the isothermal sphere, in which galaxies have reached their
equilibrium distribution. The advantage of such a model is that the volume
density is obtained by an analytical deprojection. However, de
Vaucouleurs profiles, which are much steeper in the cluster core, provide a 
better
approximation to the real distribution (Rhee \& Latour 1991~; Cirimele
et al. 1997), at the
same time leading to a finite total number of galaxies~:
\begin{equation}
\label{eq:dvc}
   \sigma_{\rm gal}(r) = \sigma_0\:{\rm exp} \paro
   - \paro \frac{r}{\rv} \parf ^{\gamma} \parf.
\end{equation}
This sort of profile was deprojected using the formula~:
\begin{equation}
\label{eq:ddvc}
   \nu(r) = - \frac{1}{2 \pi r}\:\frac{d}{dr}\;\int_{r^2}^{+\infty}
   \frac{\sigma(p)}{\sqrt{p^2 - r^2}}\:dp^2,
\end{equation}
$p$ being the projected distance to the cluster centre and $r$ the true 
distance.
Because this deprojection is numerically unstable, we computed it by assuming
$\sigma(p)$ to be constant inside a grid step and then integrating analytically
the denominator. The mass to light ratio applied to all clusters
and groups (but the supposed fossil group RXJ~1340.6+4018 consisting of only one
giant elliptical galaxy, for which we used
$\mst/\lb = 8.5 \h \msun/\lbsun$) is
$\mst/\lb = 3.2 \h \msun/\lbsun$, obtained by
White et al. (1993) by averaging over the Coma luminosity function the $\mst/L$ ratio
from van der Marel (1991) given as a function of luminosity for bright ellipticals.
Then, using the Schechter luminosity function~:
\begin{equation}
\label{eq:schech}
   n(L) dL = N^* \paro \frac{L}{L^*} \parf ^{-\alpha} e^{-\frac{L}{L^*}}\:
   d \paro \frac{L}{L^*} \parf,
\end{equation}
the luminosity emitted by a shell of thickness $dr$ and situated at the radius
$r$ writes as~: \\
$dL(r) = \ltot \times (4 \pi r^2\:\nu_{\rm gal}(r) dr)\;/\;N(> \llim)$ \\
where $N(> L) = N^*\:\Gamma(1-\alpha,L/L^*)$ is the total number of galaxies
brighter than $L$, $\llim$ being the limiting luminosity of the observations,
and $\ltot = N^* L^*\:\Gamma(2-\alpha)$. The stellar mass enclosed
in a sphere of radius $R$ can eventually be written as~:
\begin{equation}
\label{eq:stm}
   \mst(R) = \frac{\mst}{L}\:\frac{L^* \Gamma(2-\alpha)}
   {\Gamma(1-\alpha,\frac{\llim}{L^*})}\;\int_0^R 4 \pi r^2\:\nu_{\rm gal}(r) dr.
\end{equation}
When no parameters for the luminosity function were found in the literature,
we adopted the standard ones (Schechter 1975)~: $\alpha = 1.25$ and
$\mvst = - 21.9 + 5\:{\rm log} \h$. \\
\hspace*{1em}
   As a few clusters observed in X-rays do not have any available spatial galaxy
distribution (or with too poor statistics), but only either a luminosity profile
or even several total luminosities given at different radii, we then assumed
a King profile and fitted the few points by the resulting integrated luminosity
profile~:
\begin{equation}
\label{eq:mlr}
   L(< R) = L_0 \crocho \lnep \paro \frac{R}{\rc} + \paro 1 +
   \paro \frac{R}{\rc} \parf ^2 \parf ^{\frac{1}{2}} \parf - \frac{R}{\rc}
   \paro 1 + \paro \frac{R}{\rc} \parf ^2 \parf ^{-\frac{1}{2}} \crochf
\end{equation}
by varying simultaneously $L_0$ and $\rc$.
In addition to those cases, RXJ~1340.6+4018 was treated in a special way~:
we deprojected a de Vaucouleurs luminosity profile (Ponman et al. 1994).

\subsection{The X-ray gas mass profile}

\hspace*{1em}
 In their pioneering work,  Cavaliere \& Fusco-Femiano (1976) have shown under the isothermality assumption, that the X-ray gas profile is described by:
\begin{equation}
   \rho_{\rm gas}(r) = \rho_0 \paro 1 + \paro \frac{r}{\rcx} \parf ^2
   \parf ^{-\frac{3}{2} \beta}
\end{equation}

which translates to the observed X-ray surface brightness with the following
 simple analytical form~ (the so-called $\beta$-model):
\begin{equation}
   S(\theta) = S_0 \paro 1 + \paro \frac{\theta}{\thetac} \parf ^2
   \parf ^{-3 \beta + \frac{1}{2}},
\end{equation}

\noindent the slope $\beta$ and the core radius $\rcx$, which are interdependant in
their adjustment to the surface brightness, being generally found to range between
0.5 and 0.8 and between 100 and 400 kpc respectively. Very often, central
regions of clusters have to be excluded from the fit, due to cooling flows
resulting in an emission excess. The gas mass can be inferred accurately from 
the knowledge of $S_0$, $\beta$ and $\thetac$. Uncertainties in the gas mass 
are small in general, as long as it is computed inside a radius at which the 
emission is detected.
The relationship between the electron number density
and the gas mass density used here is $\rho_0 = 1.136\:\mprot\:\nez$
(assuming a helium mass fraction of 24\% and neglecting metals).

\subsection{The binding mass profile}

\hspace*{1em}
   Mass estimation is certainly the most critical aspect of recent studies of the
baryonic fraction in clusters. Clarifying this issue is one important 
aspect of this paper.
We derived the gravitational mass in two ways~: \vspace{1ex} \\

\noindent
$\bullet$ {\it The hydrostatic isothermal \bmodel} : \vspace{1ex} \\

First, we used the standard
IHE assumption which, using spherical symmetry,
translates into the mass profile~:
\begin{eqnarray}
   \mtot(r)&=&-\frac{k}{G \mu \mprot}\:\tx\:r\:\frac{d \lnep\:\rho_{\rm gas}(r)}{d \lnep\:r} \nonumber \\
   &=&\frac{3k}{G \mu \mprot}\:\beta\:\tx\:r \paro 1 +
      \paro \frac{r}{\rcx} \parf ^{-2} \parf ^{-1}.
\end{eqnarray}
The total mass thus depends linearly on both $\beta$ and $\tx$.
Hence, if the slope of the gas density is poorly determined (and this is the case
if the instrumental sensitivity is too low to achieve a good signal to noise ratio
in the outer parts of the cluster), it will have a drastic influence on the
derived mass.
This mass profile results in the density profile~:
\begin{equation}
   \rho_{\rm tot}(r) \propto \frac{1}{r^2} \crocho 3 \paro 1 + \paro
   \frac{r}{\rcx} \parf ^{-2} \parf ^{-1} - 2 \paro 1 + \paro
   \frac{r}{\rcx} \parf ^{-2} \parf ^{-2} \crochf
\end{equation}
and in a flat density at the cluster centre. \\
 The isothermality assumption can be questioned since Markevitch et al. (1998) found evidence for strong temperature gradients in clusters, which may lead to IHE mass estimates smaller by 30\% (Markevitch  1998). Although the
reality of these gradients has recently been questioned (Irwin et al.,  1999; White 2000).
\vspace{1ex} \\

\noindent
$\bullet$ {\it The universal density profile} : \vspace{1ex} \\
An alternative approach is to use the universal
dark matter density profile of Navarro et al. (1995, hereafter NFW) derived from their numerical simulations~:
\begin{equation}
   \frac{\rho_{\rm dark}(r)}{\rho_{\rm c}} = 1500\:\frac{r_{200}^3}{r\:(5 r + r_{200})^2}
\end{equation}
where $r_{200}$ stands for the radius from the cluster center where the mean
enclosed overdensity equals 200 (this is the virial radius) and $\rho_{\rm c}$ is the
critical density. It varies as $r^{-1}$ near the centre, being thus much steeper
than in the hydrostatic case~; NFW claim this behaviour fits
their high resolution simulations better than a flat profile. Furthermore, contrary to the \bmodel, the dark matter density profile obtained
by NFW is independent on the shape of the gas density distribution.
 This will introduce a further difference. The normalisation of the scaling laws ensures a relationship between temperature, virial radius and virial mass.  Here, this normalisation is taken from numerical simulations.
Different values have been published in the literature (see for instance Evrard, 1997~; Evrard et al., 1996, EMN hereafter~; Pen, 1998~; Bryan and Norman, 1998, BN hereafter). Frenk et al. (1999) investigated the formation of the same cluster with various hydrodynamical  numerical simulations. They found a small dispersion in the mass-temperature relation : the rms scatter $\sigma$ is found to be of $\approx 5\%$, EMN and BN lying at the edges of the values found, representing a 4 $\sigma$ difference.
EMN
  provide a scaling law between $r_{500}$ and $\tx$~:
\begin{equation}
   r_{500} = 2.48 \paro \frac{\tx}{10 \kev}
   \parf ^{\frac{1}{2}} \h^{-1}\:(1 + z)^{-\frac{1}{2}} \mpc
 \end{equation}
(in terms of comoving radius)
 which was used here to compute $r_{200}$, writing~:
\begin{eqnarray}
   \delta(r_{500}) = 500 &=& \paro \frac{4}{3} \pi r_{500}^3 \parf ^{-1}
      \int_0^{r_{500}} 4 \pi r^2\:\frac{\rho(r)}{\rho_{\rm c}}\:dr \nonumber \\
   &=& 180 f\:X^{-3} \crocho \lnep (1 + 5 X) - \frac{5X}{1 + 5 X} \crochf
\end{eqnarray}
with $X = r_{500}/r_{200}$, where $f = 1739/1500$ is a corrective factor
to transform dark matter mass into total mass, so that $\delta(r_{200})$ is
really equal to 200. Solving this equation gives $X = 0.66$.
The relation at $z = 0$ between virial mass and temperature can then be written as~:
\begin{eqnarray}
\tx = 4.73\:M_{15}(r_{200})^{\frac{2}{3}} \kev.
\end{eqnarray}
BN did provide the following constant of normalisation~:
\begin{eqnarray}
\tx = 3.84\:M_{15}(r_{200})^{\frac{2}{3}} \kev.
\end{eqnarray}
This difference is quite significant~: it does correspond to a virial mass
40\% higher. Using this normalisation will obviously significantly change the inferred gas fraction.

\section{Results}

\begin{figure}[!ht]
\includegraphics[scale=0.3,angle=-90]{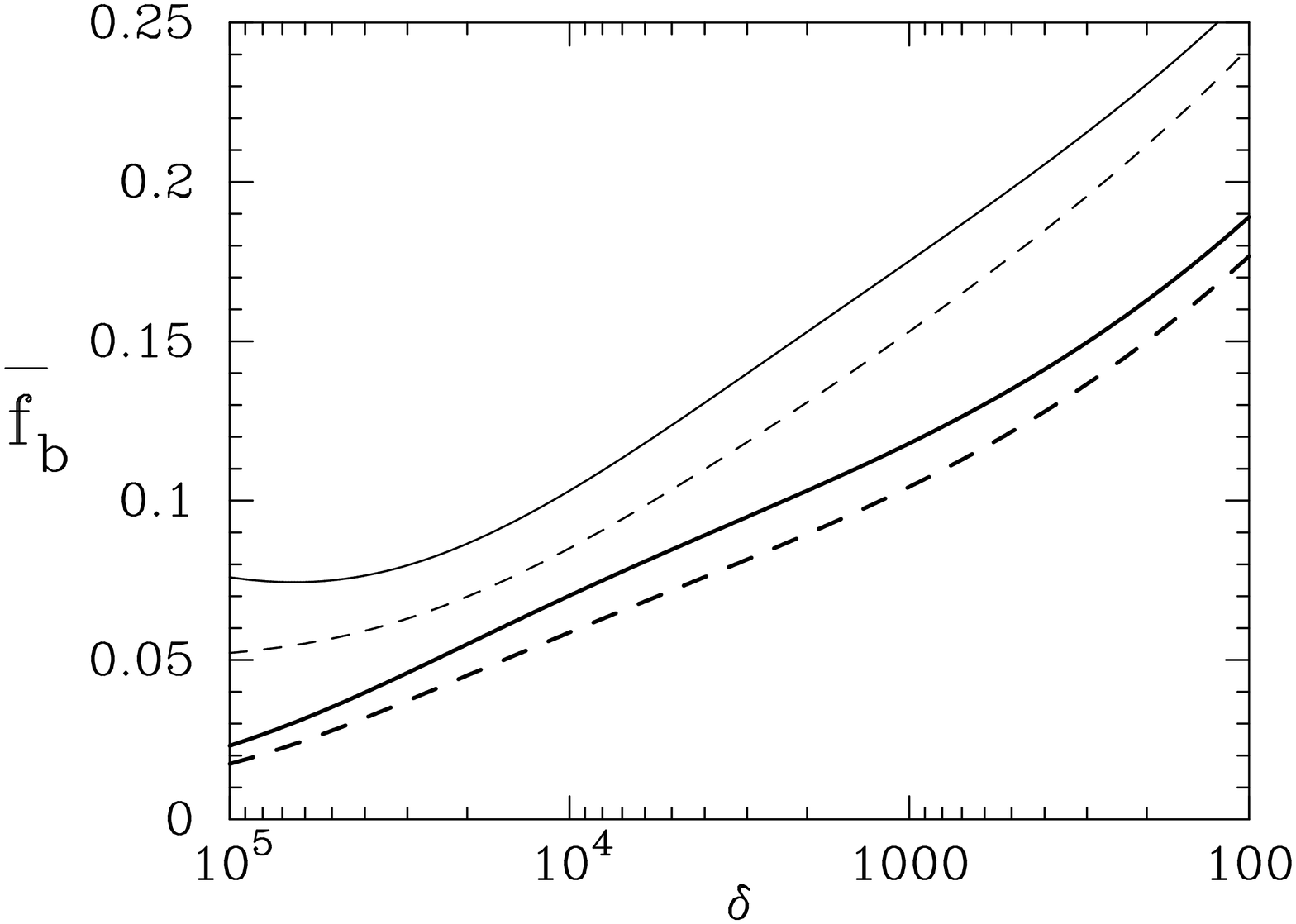}
\includegraphics[scale=0.3,angle=-90]{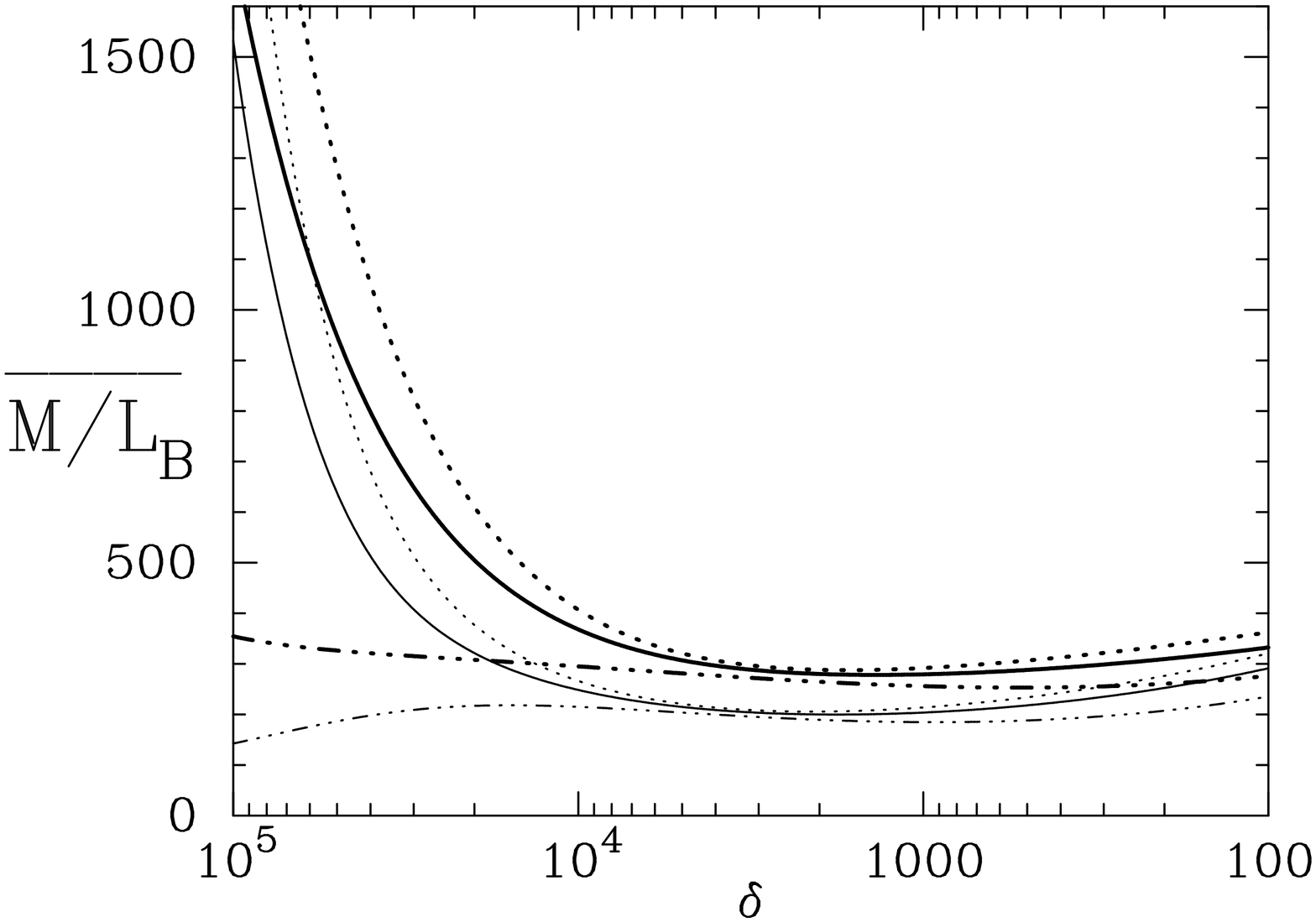}
\includegraphics[scale=0.3,angle=-90]{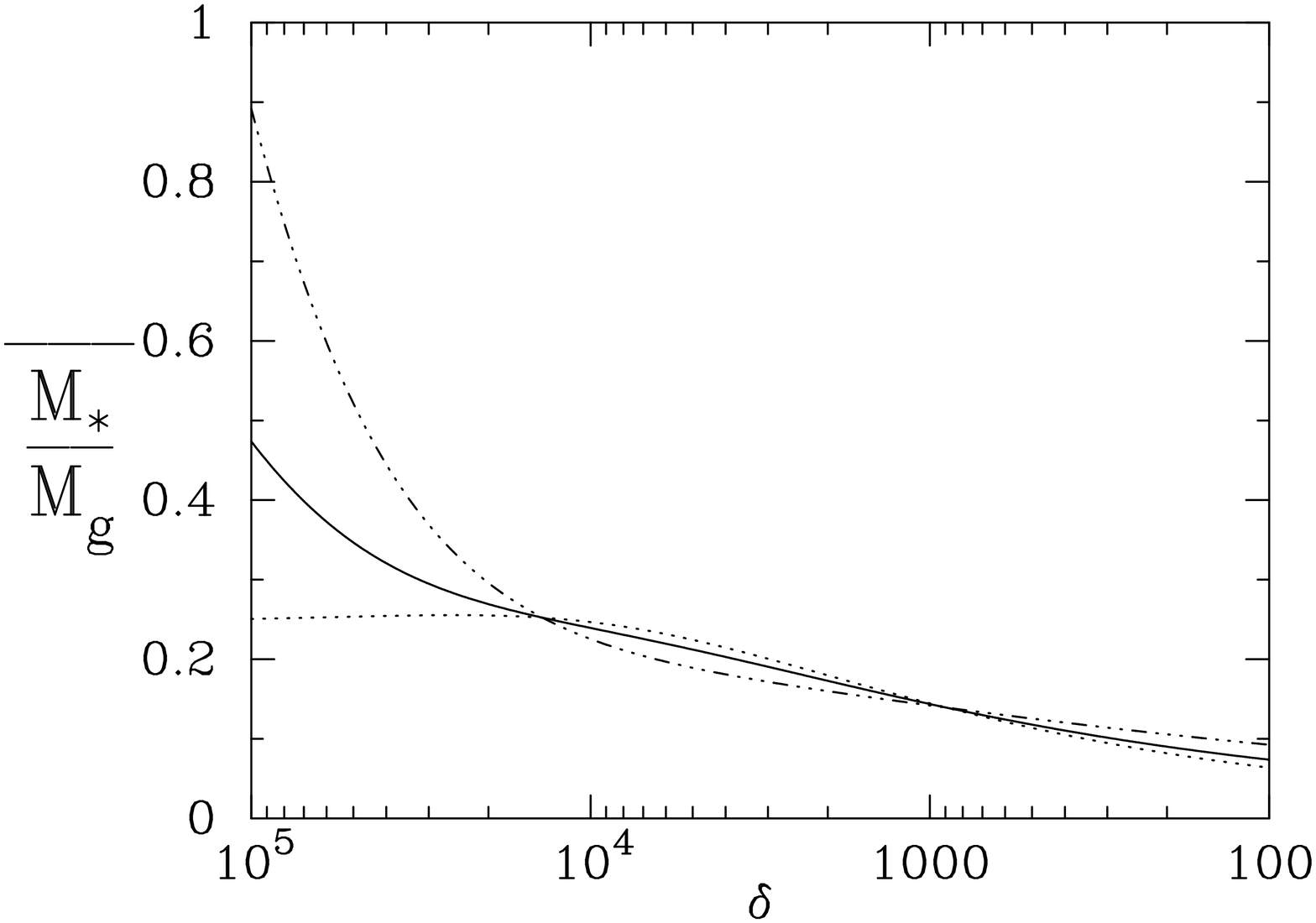}
\caption[]{Average profiles for all clusters (groups included) with the most
           reliable data. 
           Top~: baryon fraction (continuous lines) and gas fraction
           (dashes) in the case of SLM mass estimates with the EMN normalisation
           (thick lines) and hydrostatic masses (thin lines).
           Middle~: mass to luminosity
           ratio for the whole sample (continuous lines), for King galaxy
           profiles only (dots) and for de Vaucouleurs profiles only (dash-dots),
           with the same convention as previously. Bottom~: stellar mass
           to gas mass ratio, with the same line styles as for $\mlb$.
\label{fig:mprofiles}}
\end{figure}

\subsection{Distribution of the various components}

\hspace*{1em}
   The results from our analysis for each cluster in our sample are
summarised in Tables 3 to 6~: in Table 3 and Table 5 mass estimates at $r_{200}$
are derived from the SLM with the \txmv calibrations
respectively given by EMN and BN (used to compute $r_{200}$).
In Table 4 and Table 6 are summarised mean dynamical quantities over the sample at three different overdensities, $r_{200}$, $r_{500}$ and $r_{2000}$. The same
quantities are also given with mass estimates from the IHE model.

\subsubsection{The binding mass}

\hspace*{1em}
   Mass profiles of the various components for a few clusters are displayed 
in figure \ref{fig:profiles}, together with mass
ratio profiles (right side). Figure \ref{fig:tprofiles} shows baryon and gas fraction profiles for the whole sample.
Quantities are plotted against the mean enclosed contrast
density, which is the natural variable in the scaling model.
A clear feature arising from
Figure \ref{fig:profiles} concerns the different behaviours of hydrostatic masses and
total masses deduced from NFW's dark matter profile, normalised by the EMN
\txmv relationship~: NFW profiles are more centrally concentrated,
as could be foreseen from Equ.11 and 12, a property which is in agreement with the 
density profile of clusters inferred from lensing (Hammer 1991, Tyson et al. 1990).
In the outer part, when the contrast density is smaller than a few $10^4$, the shapes
of the density profiles are quite similar, although some difference in the 
amplitude exists. In fact, profiles calibrated from the EMN \txmv
relation tend to be systematically more massive than with the isothermal hydrostatic
model, with a significant dispersion. The last column of Tables 3 and 5 gives the 
ratio between masses computed with both methods. The mean of masses estimated
by the IHE \bmodel is significantly smaller than SLM masses (at 
$\rxl$): $M_{IHE}/M_{SLM} = 0.80 \pm 0.03$ with EMN normalisation, and $M_{IHE}/M_{SLM} = 0.67 \pm 0.026$ with BN normalisation.
Clearly, such a difference will translate into the baryon fraction 
estimates.

\subsubsection {The X-ray gas}

\begin{figure*}[!ht]
\hspace*{-0.5cm}
\begin{minipage}[t]{9cm}
\includegraphics[scale=0.35,angle=-90]{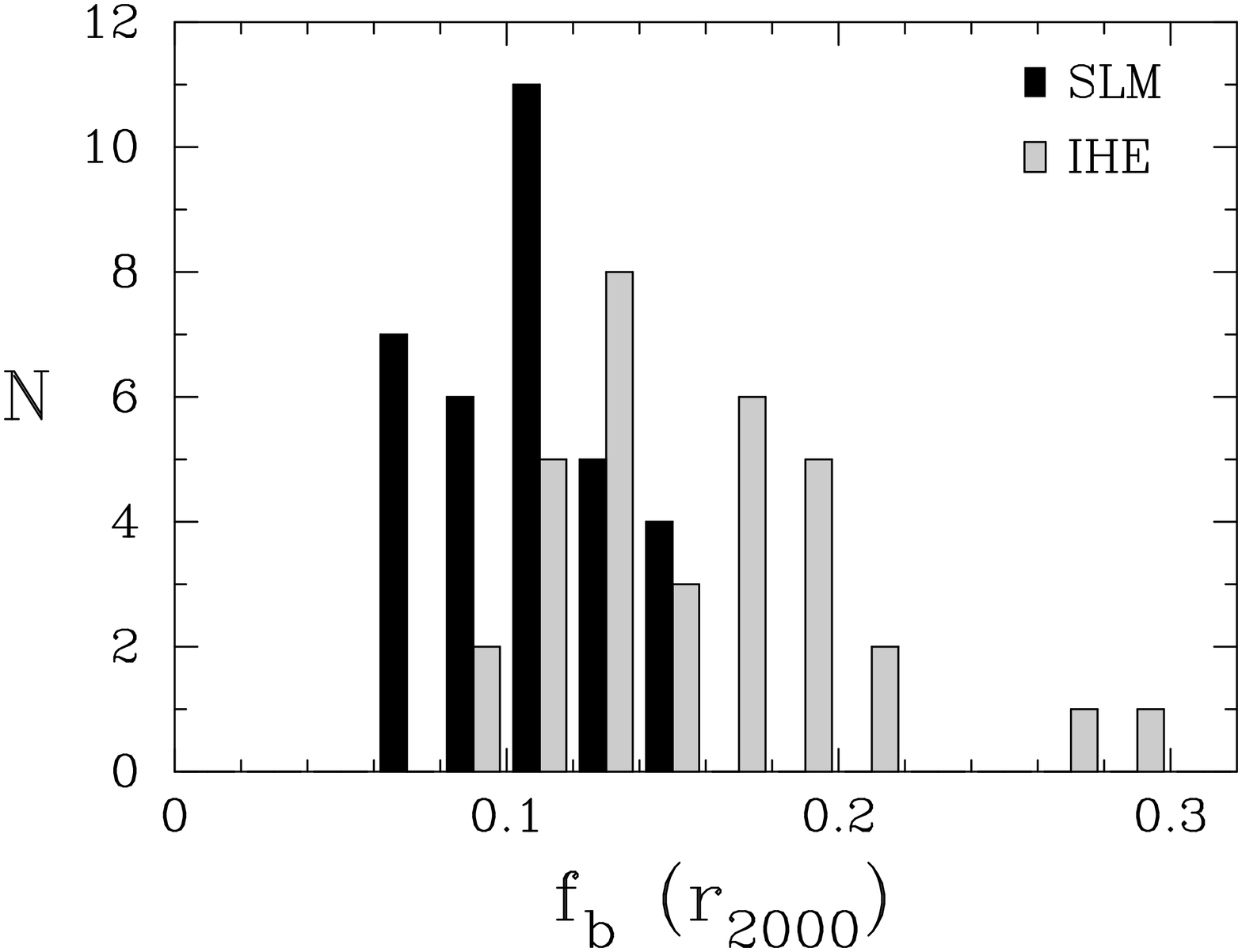}
\end{minipage}
\begin{minipage}[t]{9cm}
\includegraphics[scale=0.35,angle=-90]{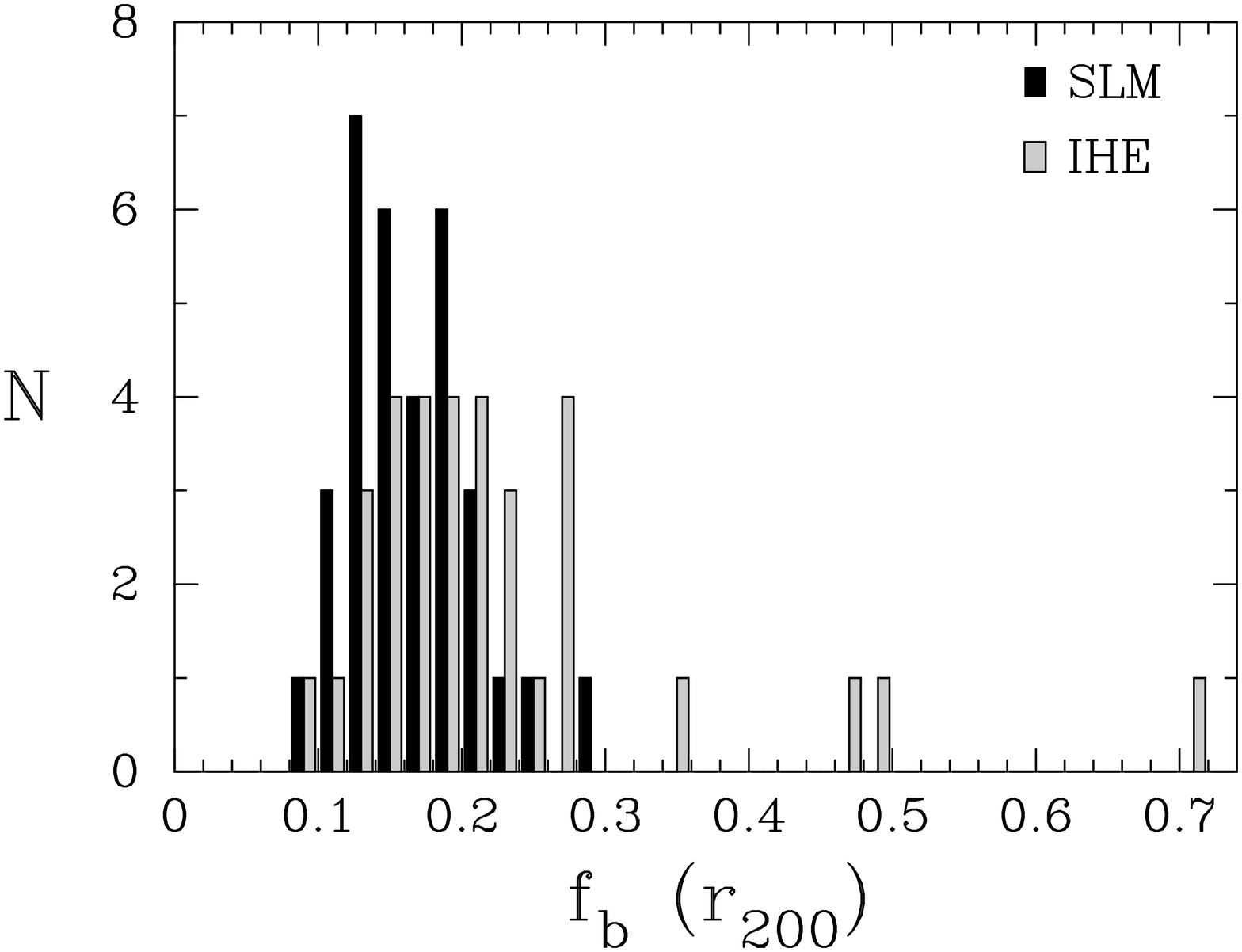}
\end{minipage}
\caption[]{Histogram of baryon fractions at $r_{2000}$ and $r_{200}$, with 
           IHE masses in grey and SLM masses in black.
           The object at $\fb = 0.7$ is NGC 4261, which has the lowest
           X-ray slope $\beta = 0.31$.\label{fig:histo}}
\end{figure*}

   Second, the distribution of gas is more spread out than that of dark matter,
which results in steadily 
rising baryon fractions with radius (Fig. \ref{fig:mprofiles}), as was already
pointed out by numerous teams, among which Durret et al. (1994) and D95. NFW also
recover this trend in their simulations. This fact makes the choice of the
limiting radius an important matter. In particular, extrapolating
masses to the virial radius (which is reached by the gas emission in only
five clusters among our sample)
could be very unsafe, especially for cool clusters, the gas of the most extended
of our objects with $\tx \leq 5 \kev$ being detected only out to $\delta = 500$.

\subsubsection{Mass to light ratio}

\hspace*{1em}
   The derived mean mass to blue luminosity ratio is shown in figure \ref{fig:mprofiles}. As 
it can be seen, $\mlb$ remains remarkably constant from $\delta \simeq 
5000$ to the outer parts of
clusters, in the case of total masses derived from SLM as
well as that of hydrostatic masses. Thus, the widely spread assumption that 
light traces mass is confirmed, at least at $r \geq \rc$.
The influence of the choice of de Vaucouleurs galaxy density profiles as compared
to King profiles is also clearly highlighted. In fact, in the core, dark matter
is normally much more concentrated than galaxies, but using a de Vaucouleurs
distribution, it turns out that the concentration factor is considerably lowered
and even reversed in the case of hydrostatic masses. Mixing the two shapes of
galaxy distribution in our sample, the result is an intermediate behaviour.

\subsection{The baryon fraction}

\hspace*{1em}
We find that inside a same object, the gas and baryon fractions increase from the center to outer shells (Fig. \ref{fig:tprofiles} and Fig. \ref{fig:mprofiles}), reflecting the fact that the distribution of gas is flatter than
that of dark matter, a trend similar to what is found by D95. Secondly, an interesting feature can be noted from figure \ref{fig:tprofiles}: the baryon fraction profiles versus density contrast are remarkably similar and seem to follow a regular behaviour, consistent with   an universal  baryon fraction  shape, even in the central part (although with a larger dispersion). This behaviour appears more clearly when one is using the SLM model. This result is consistent with the baryon fraction following a scaling law as it has been already found for the emissivity profiles (Neumann \& Arnaud, 1999)  and gas profiles (Vikhlinin et al., 1999). 
Thirdly, the comparison of the graphs of figure \ref{fig:tprofiles} shows that the baryon fraction $f_{SLM}$ estimated from 
the NFW profile normalised with the EMN \txmv relationship is
less dispersed at all 
contrast densities. This effect is asymmetric~: the high baryon fractions $f_{IHE}$
found with the IHE method disappear.  The fact that $f_{SLM}$ appears 
less dispersed has already been found by 
Evrard (1997) and Arnaud \& Evrard (1999). However, our work indicates that 
this feature exists at any radius. We also plot in Figure \ref{fig:histo} the histograms of baryon fractions as derived from both 
the IHE and SLM methods at the virial radius $r_{200}$ but also at $r_{2000}$ chosen because each object of the sample is detected in X-rays 
at least out to $\delta \sim 2000$. The comparison of the two
indeed provides an evidence for SLM masses to lead to more 
tightened baryon fractions than hydrostatic masses.  At the virial radius,
 we found that the intrinsic dispersion is 50\% with the IHE and 20\% with SLM.
This bears an important consequence
for the interpretation of mass estimates as well as the interpretation of 
the baryon fraction. Clearly the fact that the baryon fraction is less 
dispersed in the SLM even in the central regions shows that this mass estimate is safer and that the IHE method provides less 
accurate mass estimates even in the central region where hydrostatic 
equilibrium is expected to hold.

 \subsubsection{Stellar to gas mass ratio $\mst/\mgas$ }

\hspace*{1em}
   Also shown in figure \ref{fig:mprofiles} is the mean $\mst/\mgas$ ratio as a function of
overdensity, slowly going down after the central part. The galaxy density is
indeed steeper than that of gas, decreasing in $r^{-3}$ with $\epsilon = 1$
instead of $r^{-2}$ for a typical value of $\beta = 0.66$, and the situation is
even worse when a de Vaucouleurs profile is used for the galaxy distribution. 
Again, the latter contributes in a large amount to the steep decrease in the
central regions, whereas there the ratio is flat with King profiles.

\subsection{Numerical results}

\begin{table*}[!t]
\centering
\caption[]{Dynamical quantities for the whole sample, at the limiting radius
           $r_{200}$ and with the $r_{500}$--$\tx$ normalization of EMN
	   (the redshift is taken into account in this relation).
	   The scalings with the Hubble constant are~:
	   $\mtot$ and $\mst \propto \h^{-1}$,
	   $\mgas \propto \h^{-\frac{5}{2}}$ and $L \propto \h^{-2}$.}
\begin{tabular}{|c|ccccccc|}
\noalign{\smallskip}
\hline
\noalign{\smallskip}
name & $r_{200}$ & $\mtot$ & $\mgas$ & $\mst$ & $\fb$ & $\mlb$ & $M_{\rm IHE}/M_{\rm SLM}$ \\
   ~ & (Mpc) & \multicolumn{3}{c}{($10^{13} \msun$)} & (\%) & ($\msun/\lbsun$) & ~ \\
\noalign{\smallskip}
\hline
\noalign{\smallskip}
A76      &   1.37 &    20.12 &     5.42 &     0.27 &   28.3 &   239 &   0.61 \\
A85      &   2.72 &   129.70 &    10.21 &     2.90 &   10.1 &   143 &   0.70 \\
A119     &   2.64 &   125.44 &    19.21 &     1.97 &   16.9 &   204 &   0.72 \\
A401     &   3.07 &   225.02 &    44.32 &     1.56 &   20.4 &   462 &   0.75 \\
A426     &   3.01 &   177.79 &    33.23 &     0.81 &   19.2 &   700 &   0.91 \\
A576     &   2.25 &    75.29 &    10.74 &     0.99 &   15.6 &   243 &   0.84 \\
A665     &   2.65 &   183.81 &    27.92 &     1.60 &   16.1 &   367 &   0.98 \\
A1060    &   2.05 &    49.42 &     4.39 &     0.24 &    9.4 &   663 &   0.87 \\
A1377    &   1.77 &    37.30 &     3.87 &     0.96 &   13.0 &   124 &   0.62 \\
A1413    &   2.63 &   158.22 &    21.86 &     1.01 &   14.5 &   499 &   0.83 \\
A1656    &   3.26 &   221.40 &    33.66 &     1.46 &   15.9 &   486 &   0.98 \\
A1689    &   2.81 &   223.32 &    38.49 &     2.43 &   18.3 &   294 &   0.92 \\
A1775    &   2.06 &    62.65 &     8.33 &     0.83 &   14.6 &   242 &   0.78 \\
A2029    &   3.13 &   230.00 &    36.53 &     1.72 &   16.6 &   427 &   0.89 \\
A2052    &   1.96 &    48.05 &     5.38 &     0.84 &   13.0 &   183 &   0.91 \\
A2063    &   2.12 &    61.68 &     8.43 &     0.79 &   15.0 &   250 &   0.67 \\
A2163    &   3.44 &   443.64 &    82.83 &     6.97 &   20.2 &   204 &   0.77 \\
A2199    &   2.34 &    80.81 &     9.24 &     1.13 &   12.8 &   228 &   0.85 \\
A2218    &   2.44 &   144.83 &    24.80 &     2.78 &   19.0 &   167 &   0.89 \\
A2256    &   2.88 &   179.35 &    35.88 &     2.02 &   21.1 &   285 &   0.95 \\
A2634    &   2.05 &    55.71 &     7.14 &     1.43 &   15.4 &   125 &   0.75 \\
A2657    &   2.18 &    71.43 &    12.26 &     0.76 &   18.2 &   303 &   0.67 \\
A2670    &   2.06 &    61.11 &     6.16 &     0.96 &   11.6 &   204 &   0.97 \\
AWM7     &   2.25 &    73.61 &    13.05 &     0.56 &   18.5 &   423 &   0.68 \\
Hydra A  &   2.07 &    59.78 &     7.38 &     0.46 &   13.1 &   420 &   0.95 \\
Fornax   &   1.23 &    10.83 &     1.29 &     0.10 &   12.8 &   337 &   0.78 \\
HCG62    &   1.14 &     9.34 &     1.50 &     0.15 &   17.6 &   196 &   0.50 \\
HCG94    &   1.85 &    43.90 &     7.61 &     0.51 &   18.5 &   273 &   0.74 \\
NGC533   &   1.18 &    10.01 &     1.10 &     0.15 &   12.5 &   207 &   0.91 \\
NCG2300  &   1.16 &    10.33 &     2.31 &     0.13 &   23.7 &   249 &   0.49 \\
NGC4261  &   1.08 &     8.68 &     1.92 &     0.30 &   25.6 &    93 &   0.36 \\
NGC5044  &   1.16 &     8.98 &     0.81 &     0.17 &   10.9 &   171 &   0.74 \\
RXJ      &   0.90 &     6.65 &     0.69 &     0.15 &   12.6 &   144 &   1.32 \\
\noalign{\smallskip}
\hline
\end{tabular}
\end{table*}

\begin{table*}[!t]
\centering
\caption[]{Average dynamical quantities for the objects with the most reliable data
           (details on those that have been discarded can be found in the paragraph
	   ``notes on individual clusters''), using the EMN normalization.
	   Values are given  virial radius $r_{200}$ (from SLM) but also at two other ones~:
	   $r_{500}$, the limiting radius within which EMN
	   claim that the hydrostatic equilibrium is universally reached~;
	   $r_{2000}$ which we preferred to use because this represents the maximal
	   extent of X-ray observations that is valid for the whole sample, including
	   groups.
	   Groups are defined by $\tx < 2 \kev$
	   and hot clusters by $\tx > 5 \kev$.}
\begin{tabular}{|c|c|ccccc|cccc|}
\cline{3-11}
\multicolumn{2}{c|}{~} & \multicolumn{5}{c|}{~} & \multicolumn{4}{c|}{~} \\
   \multicolumn{2}{c|}{~} & \multicolumn{5}{c|}{SLM (NFW's DM profile)} &
   \multicolumn{4}{c|}{IHE (hydrostatic equilibrium)} \\
   \multicolumn{2}{c|}{~} & \multicolumn{5}{c|}{~} & \multicolumn{4}{c|}{~} \\
\cline{3-11}
\noalign{\smallskip}
\multicolumn{2}{c|}{~} & $\fg$ & $\fb$ & $\mgas/\mst$ & $\mtot/\mst$ & $\mlb$ &
   $\fg$ & $\fb$ & $\mtot/\mst$ & $\mlb$ \\
   \multicolumn{2}{c|}{~} & (\%) & (\%) & ~ & ~ & ($\msun/\lbsun$) &
   (\%) & (\%) & ~ & ($\msun/\lbsun$) \\
\noalign{\smallskip}
\hline
\noalign{\smallskip}
            ~ & at $r_{200}$ &   14.6 &   16.0 &   13.6 &    93.3 &   298 &   20.1 &   22.0 &    77.3 &   247 \\
          all & at $r_{500}$ &   11.9 &   13.4 &   10.5 &    86.9 &   278 &   17.0 &   19.2 &    66.6 &   213 \\
           ~ & at $r_{2000}$ &    8.9 &   10.4 &    7.7 &    84.5 &   270 &   12.8 &   15.2 &    60.9 &   195 \\
\noalign{\smallskip}
\hline
\noalign{\smallskip}
            ~ & at $r_{200}$ &   14.7 &   16.5 &    9.0 &    62.4 &   200 &   26.6 &   29.8 &	 45.7 &   146 \\
       groups & at $r_{500}$ &   11.0 &   13.1 &    6.3 &    58.3 &   186 &   19.7 &   23.4 &	 39.0 &   125 \\
           ~ & at $r_{2000}$ &    7.0 &    9.6 &    4.2 &    61.1 &   196 &   12.3 &   16.6 &	 37.5 &   120 \\
\noalign{\smallskip}
\hline
\noalign{\smallskip}
	    ~ & at $r_{200}$ &   13.4 &   14.6 &   12.2 &    95.0 &   304 &   17.2 &   18.8 &    77.8 &   249 \\
cool clusters & at $r_{500}$ &   11.0 &   12.3 &    9.3 &    88.6 &   283 &   15.0 &   16.7 &    67.9 &   217 \\
	   ~ & at $r_{2000}$ &    8.4 &    9.7 &    6.9 &    84.6 &   271 &   11.6 &   13.5 &    63.2 &   202 \\
\noalign{\smallskip}
\hline
\noalign{\smallskip}
	    ~ & at $r_{200}$ &   15.9 &   17.0 &   19.4 &   118.1 &   378 &   18.8 &   20.2 &   104.3 &   334 \\
 hot clusters & at $r_{500}$ &   13.7 &   14.8 &   15.7 &   109.8 &   351 &   17.3 &   18.9 &    89.1 &   285 \\
	   ~ & at $r_{2000}$ &   10.7 &   11.9 &   11.8 &   104.8 &   335 &   14.4 &   16.0 &    78.5 &   251 \\
\noalign{\smallskip}
\hline
\end{tabular}
\end{table*}

\begin{table*}[!t]
\centering
\caption[]{Same as Table 3 but with the $r_{200}$--$\tx$ normalization of BN.}
\begin{tabular}{|c|ccccccc|}
\noalign{\smallskip}
\hline
\noalign{\smallskip}
name & $r_{200}$ & $\mtot$ & $\mgas$ & $\mst$ & $\fb$ & $\mlb$ & $M_{\rm IHE}/M_{\rm SLM}$ \\
   ~ & (Mpc) & \multicolumn{3}{c}{($10^{13} \msun$)} & (\%) & ($\msun/\lbsun$) & ~ \\
\noalign{\smallskip}
\hline
\noalign{\smallskip}
A76      &   1.52 &    26.47 &     6.40 &     0.31 &   25.4 &   272 &   0.52 \\
A85      &   3.02 &   174.74 &    11.97 &     3.11 &    8.6 &   180 &   0.58 \\
A119     &   2.93 &   167.47 &    22.42 &     2.27 &   14.7 &   236 &   0.60 \\
A401     &   3.41 &   297.72 &    50.75 &     1.65 &   17.6 &   577 &   0.63 \\
A426     &   3.34 &   234.89 &    37.20 &     0.86 &   16.2 &   879 &   0.77 \\
A576     &   2.50 &   100.32 &    12.21 &     1.07 &   13.2 &   299 &   0.70 \\
A665     &   2.95 &   244.16 &    31.16 &     1.72 &   13.5 &   455 &   0.83 \\
A1060    &   2.28 &    66.58 &     4.99 &     0.25 &    7.9 &   852 &   0.72 \\
A1377    &   1.97 &    50.07 &     4.63 &     0.99 &   11.2 &   162 &   0.52 \\
A1413    &   2.92 &   211.20 &    24.82 &     1.04 &   12.2 &   650 &   0.69 \\
A1656    &   3.62 &   294.04 &    37.42 &     1.53 &   13.2 &   615 &   0.82 \\
A1689    &   3.12 &   295.02 &    42.65 &     2.61 &   15.3 &   362 &   0.78 \\
A1775    &   2.29 &    83.81 &     9.59 &     0.96 &   12.6 &   278 &   0.64 \\
A2029    &   3.48 &   305.31 &    40.92 &     1.82 &   14.0 &   537 &   0.75 \\
A2052    &   2.18 &    64.21 &     6.04 &     0.89 &   10.8 &   230 &   0.76 \\
A2063    &   2.35 &    82.53 &     9.89 &     0.81 &   13.0 &   328 &   0.56 \\
A2163    &   3.82 &   586.39 &    94.45 &     7.42 &   17.4 &   253 &   0.65 \\
A2199    &   2.60 &   108.21 &    10.48 &     1.28 &   10.9 &   270 &   0.70 \\
A2218    &   2.71 &   191.28 &    27.75 &     2.97 &   16.1 &   206 &   0.75 \\
A2256    &   3.20 &   235.80 &    39.93 &     2.15 &   17.8 &   350 &   0.81 \\
A2634    &   2.28 &    74.50 &     8.31 &     1.64 &   13.4 &   146 &   0.63 \\
A2657    &   2.42 &    95.12 &    14.32 &     0.80 &   15.9 &   378 &   0.56 \\
A2670    &   2.28 &    81.87 &     6.87 &     1.06 &    9.7 &   248 &   0.80 \\
AWM7     &   2.50 &    97.92 &    15.18 &     0.58 &   16.1 &   542 &   0.57 \\
Hydra A  &   2.30 &    79.83 &     8.21 &     0.48 &   10.9 &   537 &   0.79 \\
Fornax   &   1.37 &    14.52 &     1.49 &     0.11 &   11.0 &   414 &   0.65 \\
HCG62    &   1.26 &    12.51 &     1.82 &     0.16 &   15.8 &   246 &   0.41 \\
HCG94    &   2.05 &    58.31 &     8.77 &     0.54 &   16.0 &   345 &   0.62 \\
NGC533   &   1.32 &    13.41 &     1.25 &     0.16 &   10.6 &   263 &   0.76 \\
NCG2300  &   1.28 &    13.73 &     2.79 &     0.14 &   21.4 &   308 &   0.41 \\
NGC4261  &   1.20 &    11.55 &     2.39 &     0.32 &   23.4 &   117 &   0.30 \\
NGC5044  &   1.29 &    12.07 &     0.94 &     0.18 &    9.3 &   216 &   0.61 \\
RXJ      &   1.00 &     8.85 &     0.74 &     0.15 &   10.0 &   194 &   1.11 \\
\noalign{\smallskip}
\hline
\end{tabular}
\end{table*}

\begin{table*}[!t]
\centering
\caption[]{Same as Table 4, but using the BN normalization.}
\begin{tabular}{|c|c|ccccc|cccc|}
\cline{3-11}
\multicolumn{2}{c|}{~} & \multicolumn{5}{c|}{~} & \multicolumn{4}{c|}{~} \\
   \multicolumn{2}{c|}{~} & \multicolumn{5}{c|}{SLM (NFW's DM profile)} &
   \multicolumn{4}{c|}{IHE (hydrostatic equilibrium)} \\
   \multicolumn{2}{c|}{~} & \multicolumn{5}{c|}{~} & \multicolumn{4}{c|}{~} \\
\cline{3-11}
\noalign{\smallskip}
\multicolumn{2}{c|}{~} & $\fg$ & $\fb$ & $\mgas/\mst$ & $\mtot/\mst$ & $\mlb$ &
   $\fg$ & $\fb$ & $\mtot/\mst$ & $\mlb$ \\
   \multicolumn{2}{c|}{~} & (\%) & (\%) & ~ & ~ & ($\msun/\lbsun$) &
   (\%) & (\%) & ~ & ($\msun/\lbsun$) \\
\noalign{\smallskip}
\hline
\noalign{\smallskip}
            ~ & at $r_{200}$ &   12.6 &   13.7 &   14.6 &   117.1 &   375 &   21.0 &   22.8 &    81.1 &   260 \\
          all & at $r_{500}$ &   10.3 &   11.5 &   11.2 &   107.5 &   344 &   17.7 &   19.8 &    68.8 &   220 \\
           ~ & at $r_{2000}$ &    7.7 &    9.0 &    8.0 &   100.8 &   323 &   13.4 &   15.8 &    60.9 &   195 \\
\noalign{\smallskip}
\hline
\noalign{\smallskip}
            ~ & at $r_{200}$ &   13.1 &   14.5 &    9.9 &    78.4 &   251 &   28.8 &   31.8 &    48.2 &   154 \\
       groups & at $r_{500}$ &    9.7 &   11.4 &    6.8 &    72.1 &   231 &   21.2 &   24.8 &    40.3 &   129 \\
           ~ & at $r_{2000}$ &    6.3 &    8.3 &    4.4 &    71.7 &   230 &   13.2 &   17.4 &    36.8 &   118 \\
\noalign{\smallskip}
\hline
\noalign{\smallskip}
            ~ & at $r_{200}$ &   11.5 &   12.5 &   13.1 &   119.7 &   383 &   17.7 &   19.3 &    81.4 &   260 \\
cool clusters & at $r_{500}$ &    9.4 &   10.5 &    9.9 &   110.0 &   352 &   15.5 &   17.3 &    69.9 &   224 \\
           ~ & at $r_{2000}$ &    7.3 &    8.4 &    7.1 &   101.7 &   325 &   12.1 &   14.0 &    63.1 &   202 \\
\noalign{\smallskip}
\hline
\noalign{\smallskip}
            ~ & at $r_{200}$ &   13.5 &   14.4 &   20.6 &   147.7 &   473 &   19.1 &   20.5 &   109.5 &   350 \\
 hot clusters & at $r_{500}$ &   11.7 &   12.6 &   16.5 &   135.5 &   434 &   17.7 &   19.2 &    92.2 &   295 \\
           ~ & at $r_{2000}$ &    9.3 &   10.3 &   12.3 &   125.2 &   401 &   14.9 &   16.5 &    79.2 &   253 \\
\noalign{\smallskip}
\hline
\end{tabular}
\end{table*}

\hspace*{1em}
   Average numerical results are presented in Tables 4 and 6. It is found that
the mean baryon fraction using the
 SLM with the 
\txmv normalization of EMN is 13.4\% and the gas fraction 
11.5\% at $r_{500}$ to be compared
with hydrostatic results~: respectively 19.2 and 17.0\%. As expected, the two methods of mass estimation lead to different baryon (gas) fractions. This difference is not negligible ($\approx 40\%$) and is mainly due as already noted, to the difference between the IHE  mass and the SLM mass. The IHE mass can be 50 to 60\% lower with respect to the SLM mass (this is the case, for instance, of the groups HCG~62, NGC~2300 and NGC~4261).This difference between $f_b^{SLM}$ and $f_b^{IHE}$, increases when using the \txmv normalization of BN (the mean baryon fraction being then 11.5 and the mean gas fraction 10.3\%).
Cirimele et al. (1997) found $\fb = 23$\% for their 12 clusters included
in our sample (and 20\% excluding A76),
instead of our result of 19\% (and 16\%) using their parameters and the same hydrostatic $\beta$-model and their limiting radius (they choose a uniform
 $R_{\rm X lim} = 1.5 \mpc$
) and of $13\%$ using the SLM method.
%and not the X-ray map radius), and of 12\% using NFW's profile.
%Nevertheless, when stopping at the same radius as them, we still find a 
%discrepancy between our value ($f_b = 18$\%) and theirs. 
The disagreement is due
 to the adopted stellar mass to light ratio ($ \mlb =$ 10.7 instead of our $3.2$ value).
%$6 M_{\odot} / L_{B \odot}$ value).
% and to gas masses (ours are in the mean
%11\% lower), but our hydrostatic masses are in excellent agreement with theirs.
%
%instead of our result of 19\% (16\% excluding A76 with 48\%), using their
%parameters, the same hydrostatic \bmodel and their limiting radius
%(they chose a uniform $\rxl = 1.5 \mpc$), and of 13\% using SLM.
%The discrepancy is due to the adopted stellar mass to light ratio (10.7
%instead of our $3.2$  $ \mlb $ value).
From the results of D95, it comes out that their 7 clusters
also have a mean baryon fraction of $f_b \approx 23\%$. Thus, this is a confirmation of
the divergence between hydrostatic \bmodel mass estimates and SLM's 
masses. From a sample of 26 clusters among which
%6 hot and 2 cool clusters
%
7 hot and 3 cool clusters
are in our sample, Arnaud \& Evrard (1999) have made a similar analysis
and derived in the frame of simulation-calibrated virial masses
a mean gas fraction at $\delta=500$ of $\simeq 14$\% in rhough agreement with our value
of 12\%.
If the comparison is restricted to hot cluster subsamples, the agreement is as 
good (they found 16~\% to be compared with our 14\%), and also at $r_{200}$. A somewhat higher gas fraction ($f_g \approx 17\%$) has been obtained recently by Mohr et al. (1999), as compared to ours, which is probably due to the difference in the normalization of the $T_X-M_V$ relationship.   \\
\hspace*{1em}
   Another output from the present study is the mean total mass to blue
luminosity ratio $\mlb \simeq 270$ at $r_{2000}$ (the hydrostatic assumption 
leading to $\mlb \simeq 200$), groups and clusters of all temperatures put
together. However, when looking in more detail at the three classes of groups
(with $\tx \leq 2 \kev$), cool clusters ($\tx \leq 5 \kev$)
and hot clusters, $\mlb$ (at $r_{2000}$) goes from 200 to 270 and
340 respectively, with similar statistics (7 groups, 10 cool clusters and 8 hot
clusters). Hence, we disagree with D95 who claim that the mass to light ratio is
roughly constant from groups to rich clusters (using the group NGC 5044 which
also belongs to our sample with $\mlb \simeq 160$,
2 cool clusters and 4 hot clusters, 3 of which are also in common with ours).
It is worth noticing that 2 of the 3 clusters in common have a low $\mlb$
in our analysis~: 150 for A85 and 170 for A2063.
%Still, even if there is
%a factor of 1.6 between groups and hot clusters, the conclusion does not hold 
%when examining \mlb out to \r500 or \r200~: in this case,
%the value for groups is close to that found for hot clusters, though it should
%be reminded that extrapolating masses for groups out to such large radii may be
%very unsafe. Furthermore, the high dispersion seen within each category
%prevents us from drawing a firm conclusion, as we think it is also the case for 
%other studies. Nevertheless,
%
This conclusion holds whatever the limiting radius~: there is a factor of 1.7,
 1.9 and 1.9 respectively between groups and hot clusters when examining
$\mlb$ out to $r_{2000}$, $r_{500}$ or $r_{200}$.\\ 
\hspace*{1em}
   As to the mean gas to stellar mass ratio $\mgas/\mst$, its values are
summarized in Tables 4 and 6. We have computed this quantity to estimate the stellar contribution to the baryon fraction and to investigate any
correlation with temperature, which will be discussed in the next section.
Let us simply mention that our value for groups at $r_{2000}$ is in good agreement
with the mean value $\simeq 5$ of Dell'Antonio et al. (1995) for 4 poor clusters,
after correcting for the different $\mst/L$ they have used.

\section{Correlations of the baryon population properties with temperature}

\begin{figure*}[!t]
\begin{minipage}[t]{9cm}
\includegraphics[scale=0.35,angle=-90]{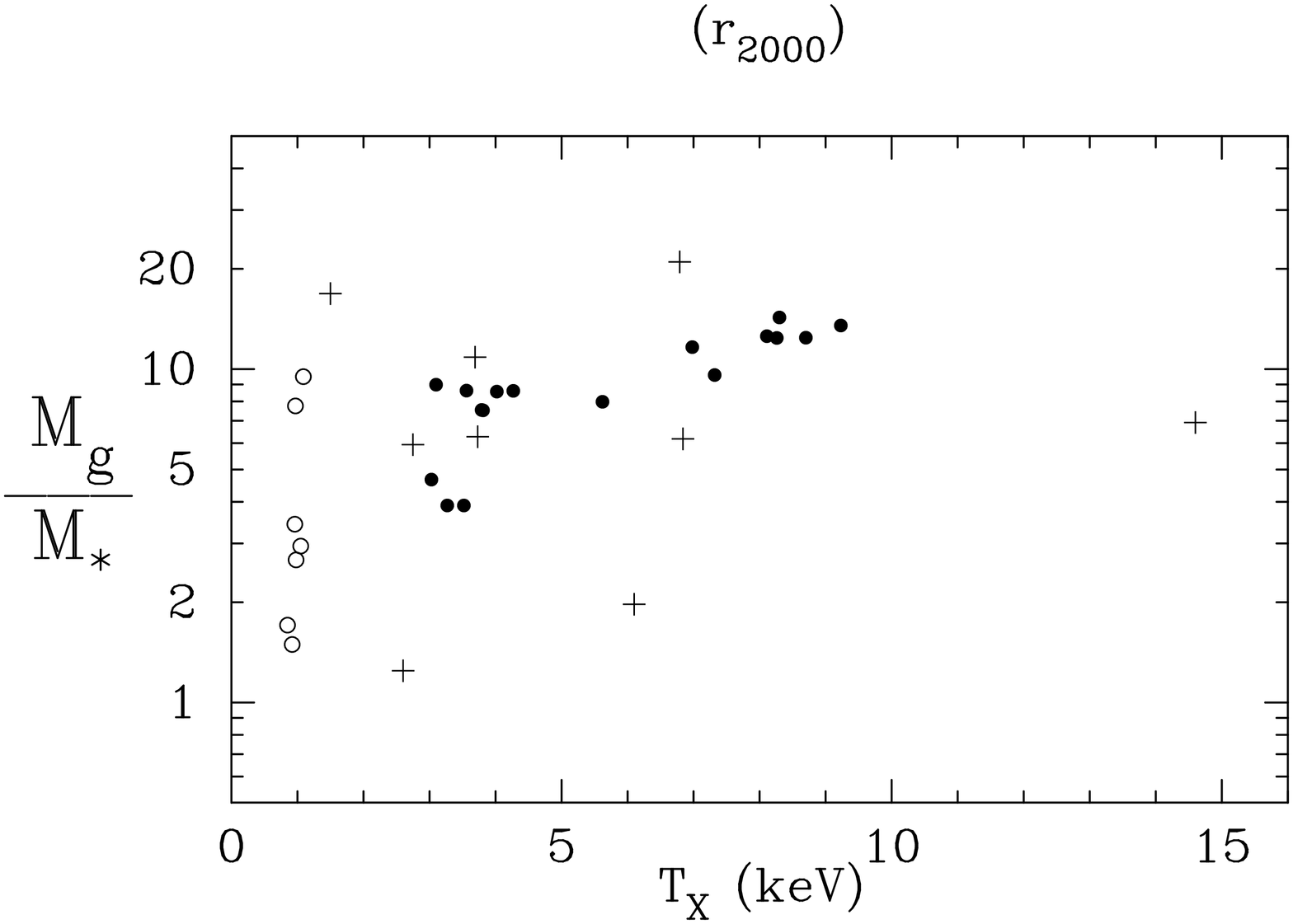}
\end{minipage}
\begin{minipage}[t]{9cm}
\includegraphics[scale=0.35,angle=-90]{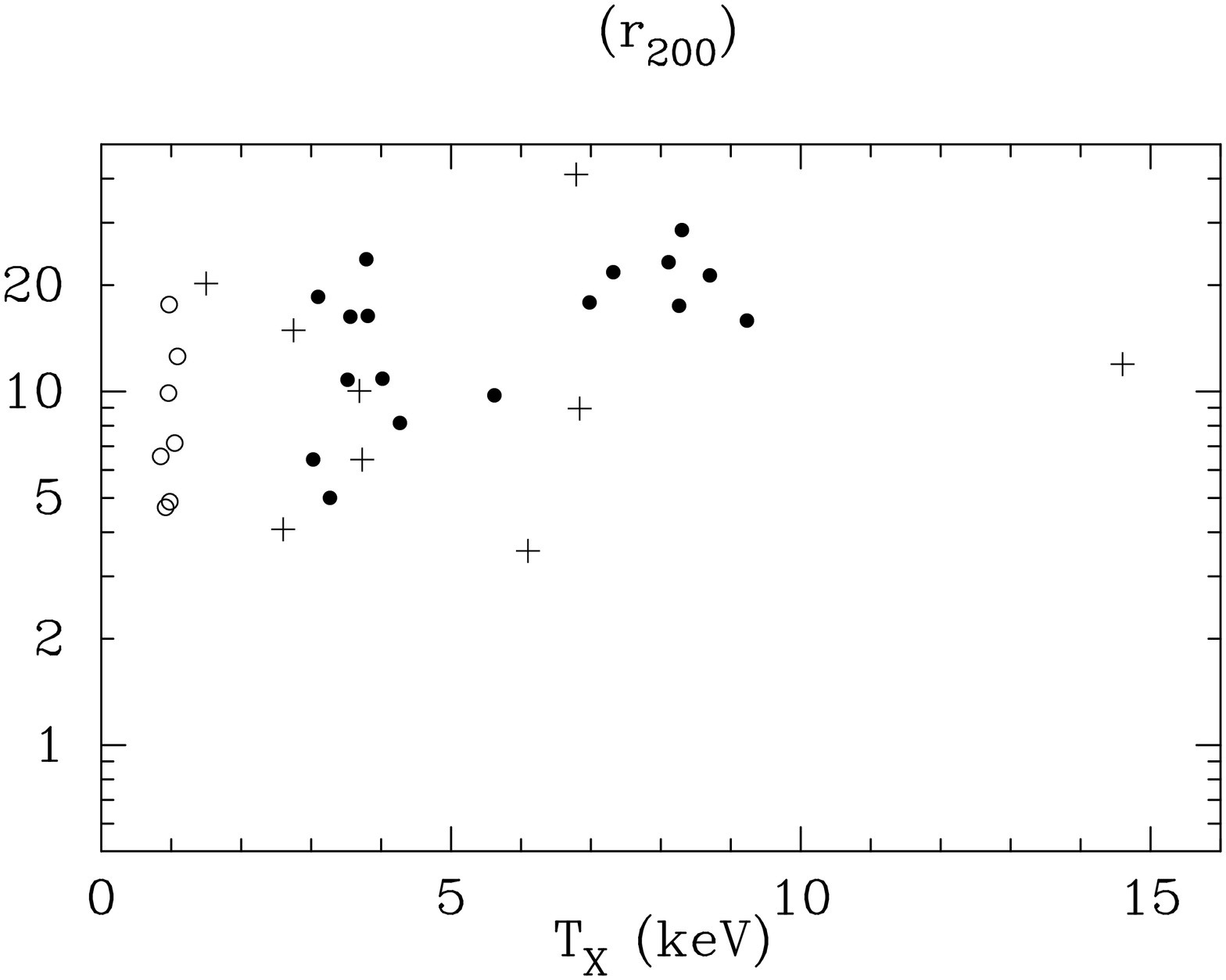}
\end{minipage}
\caption[]{Gas to galaxy mass ratio versus X-ray temperature. 
           Open circles are for groups, filled circles are for clusters, and crosses
           refer to poor quality  optical or gas masses. \label{fig:mtg}}
\end{figure*}

\begin{figure*}[!ht]
\begin{minipage}[t]{9cm}
\includegraphics[scale=0.35,angle=-90]{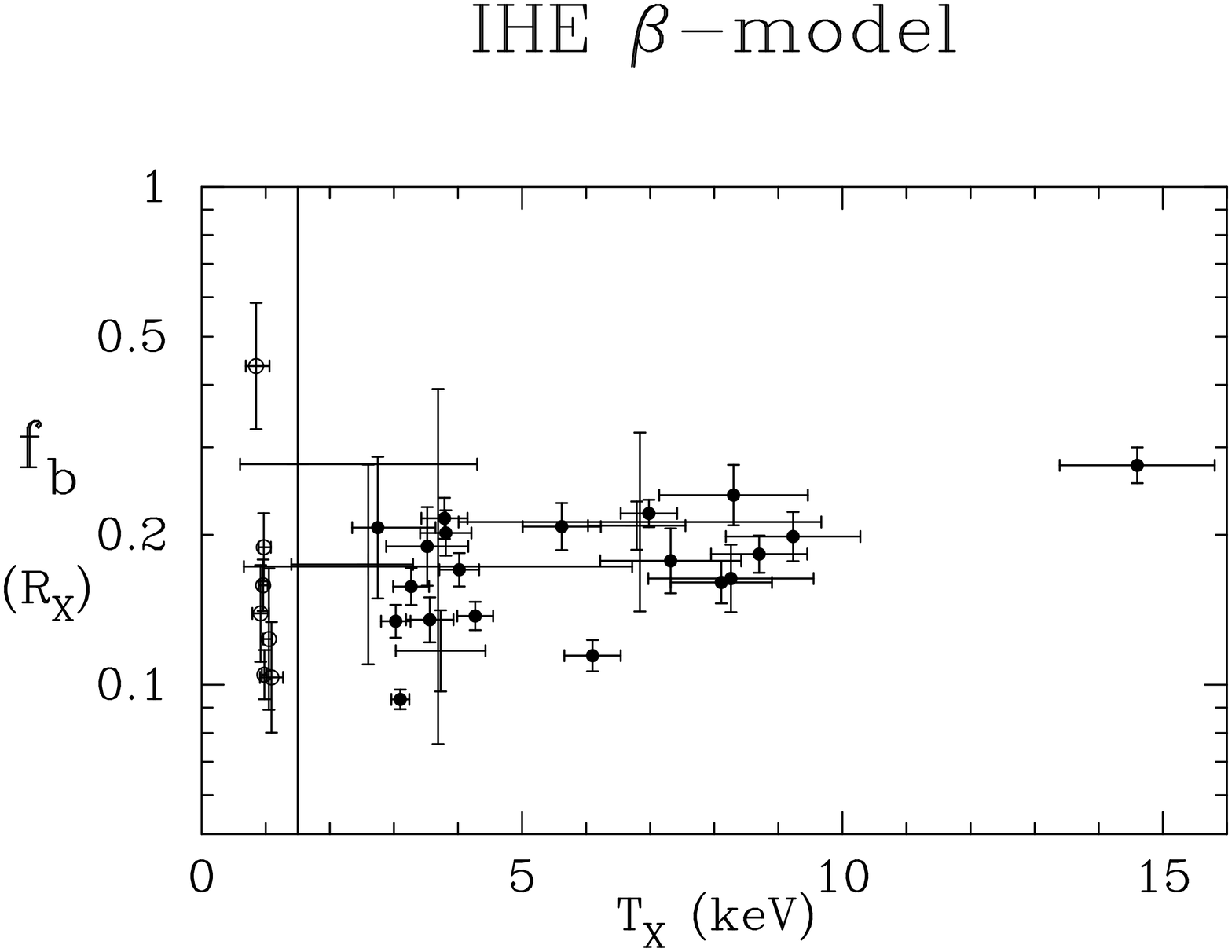}
\end{minipage}
\begin{minipage}[t]{9cm}
\includegraphics[scale=0.35,angle=-90]{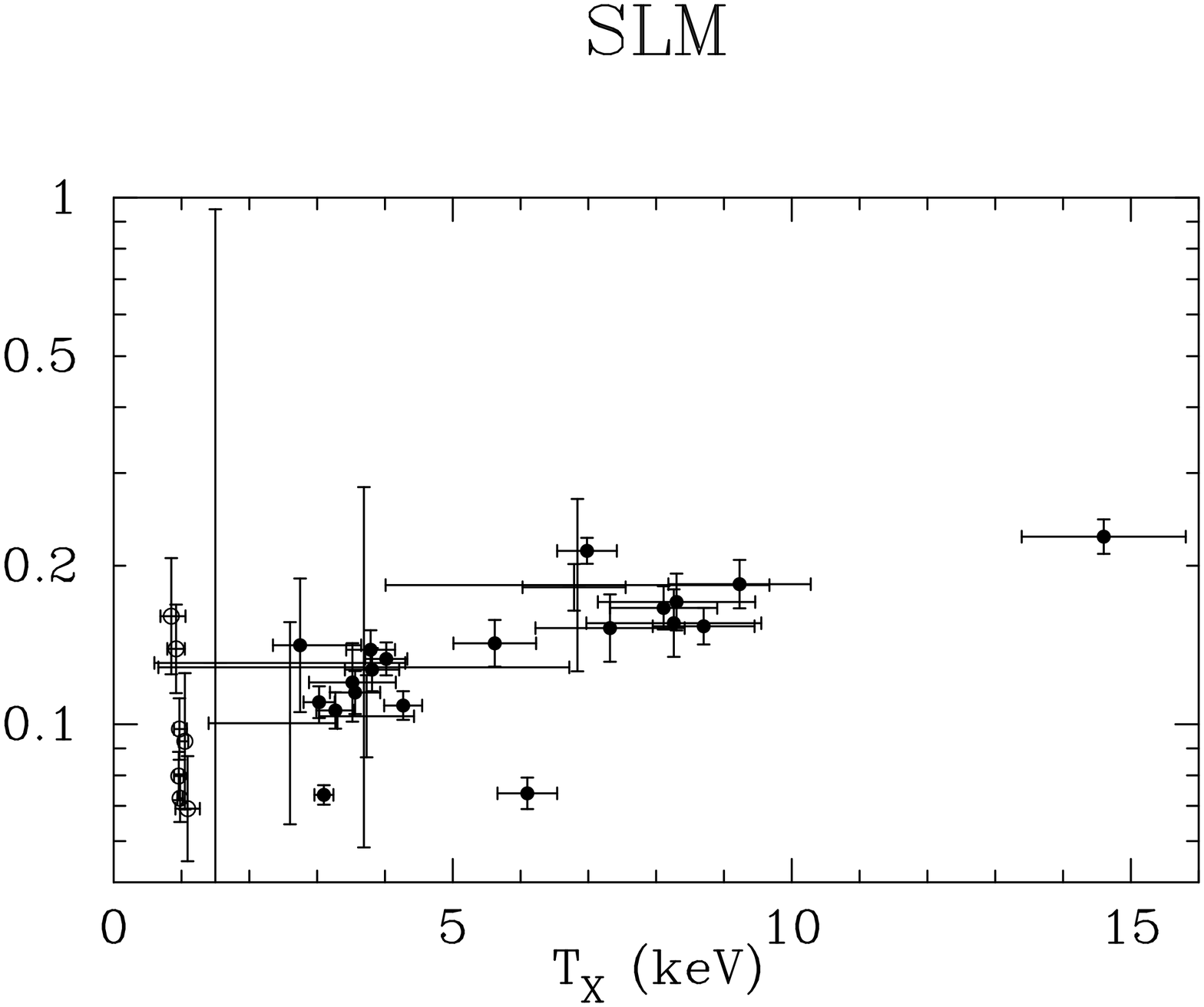}
\end{minipage}
\vspace*{-1.cm} \\
\begin{minipage}[t]{9cm}
\includegraphics[scale=0.35,angle=-90]{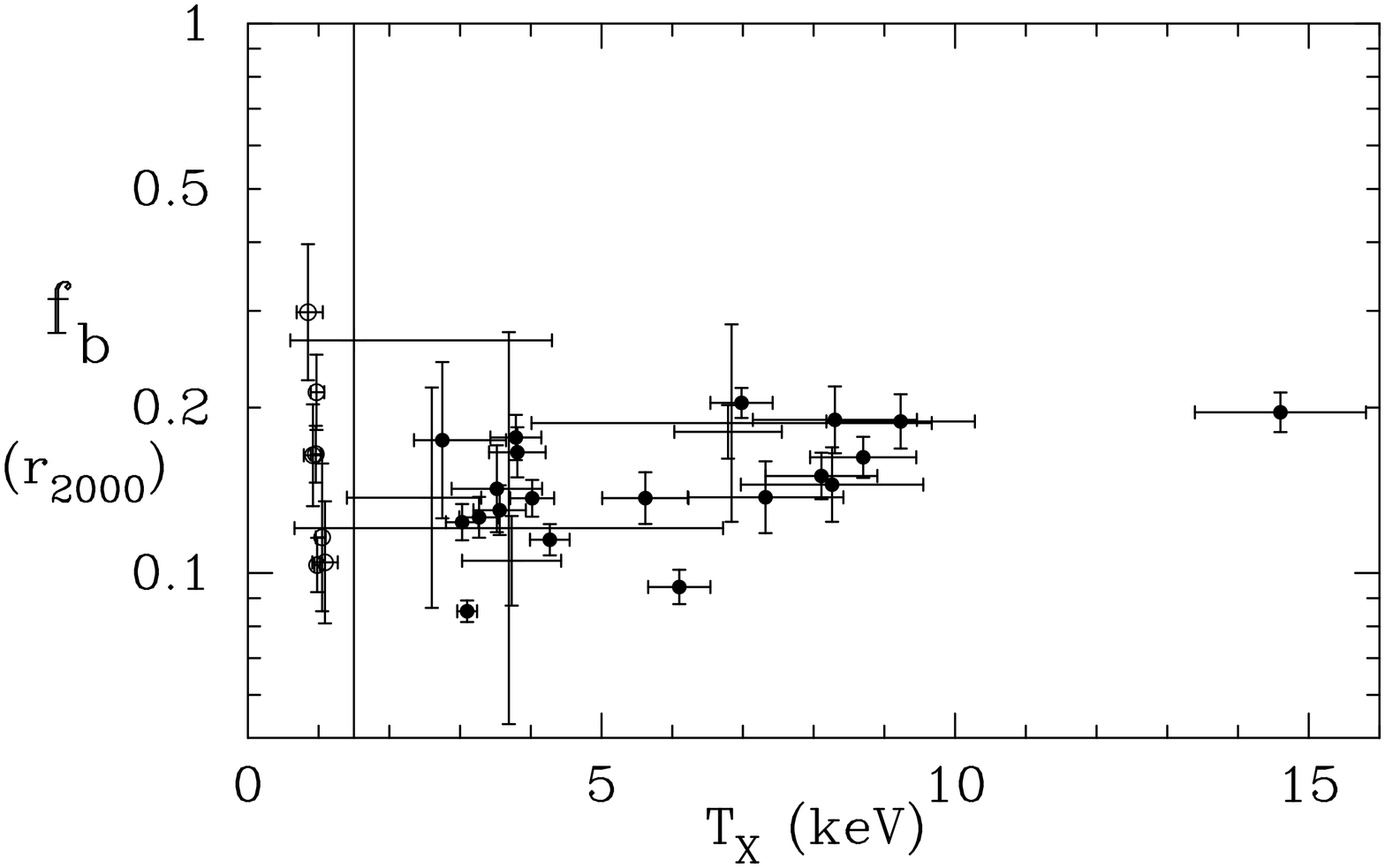}
\end{minipage}
\begin{minipage}[t]{9cm}
\includegraphics[scale=0.35,angle=-90]{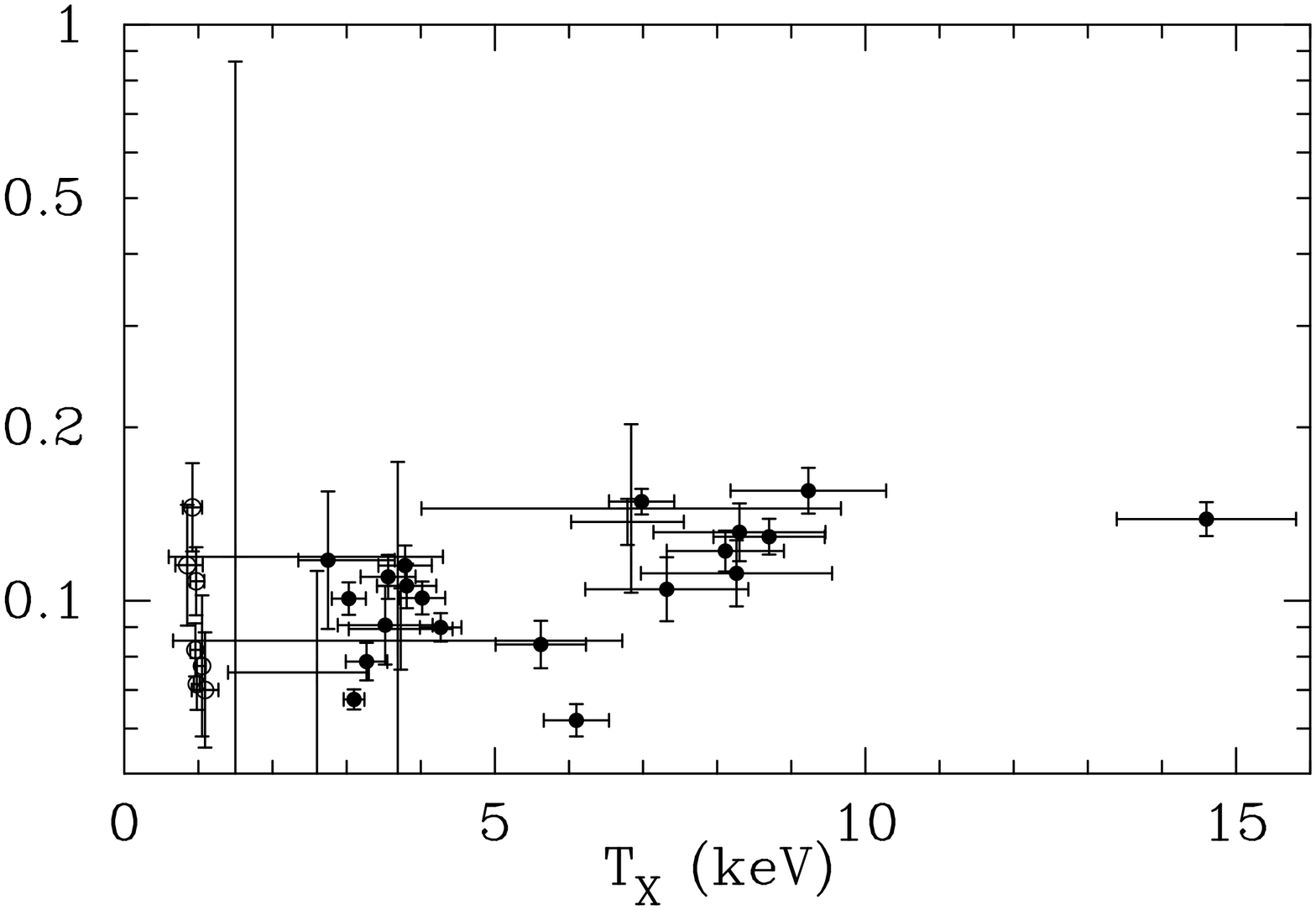}
\end{minipage}
\vspace*{-1.cm} \\
\begin{minipage}[t]{9cm}
\includegraphics[scale=0.35,angle=-90]{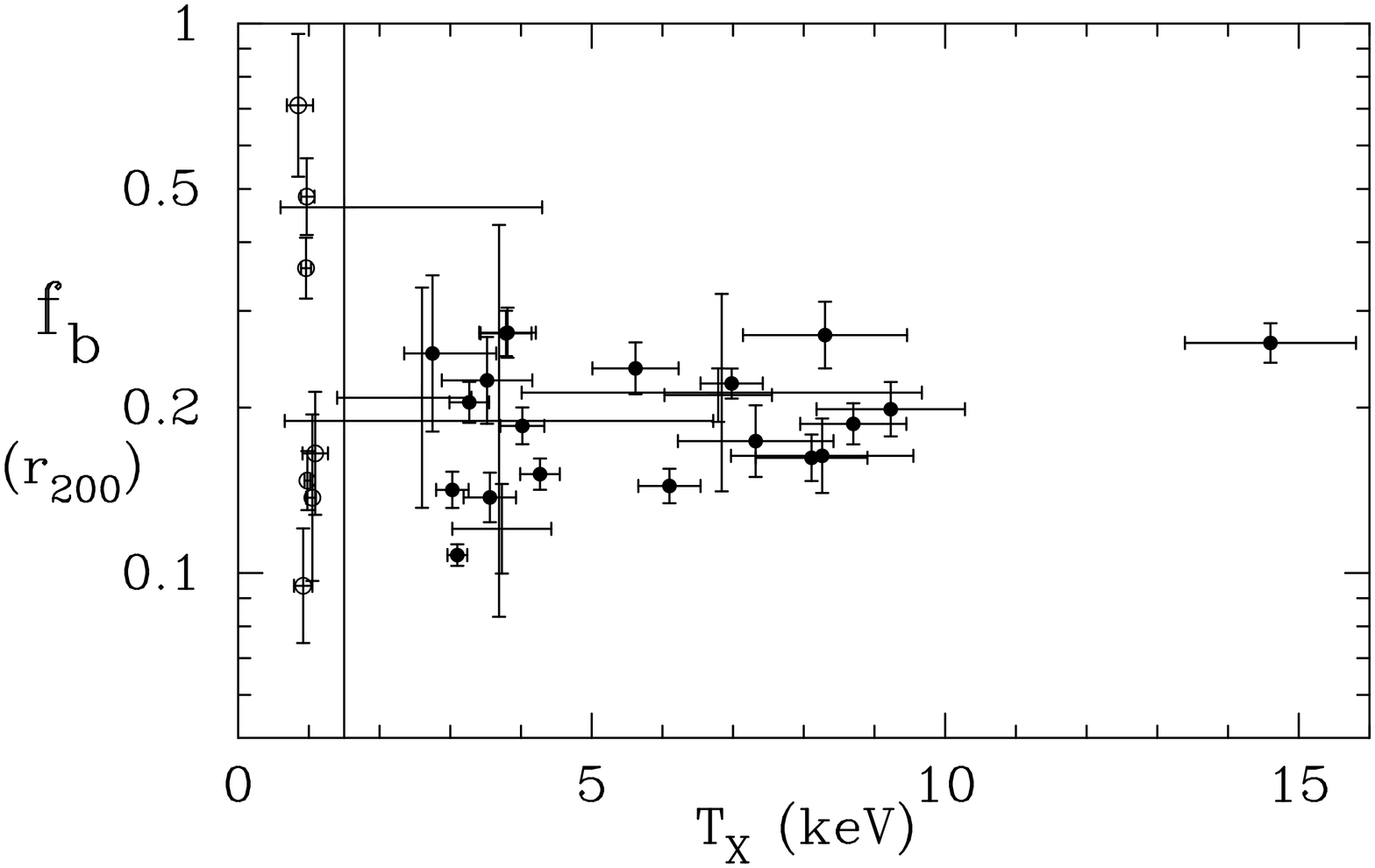}
\end{minipage}
\begin{minipage}[t]{9cm}
\includegraphics[scale=0.35,angle=-90]{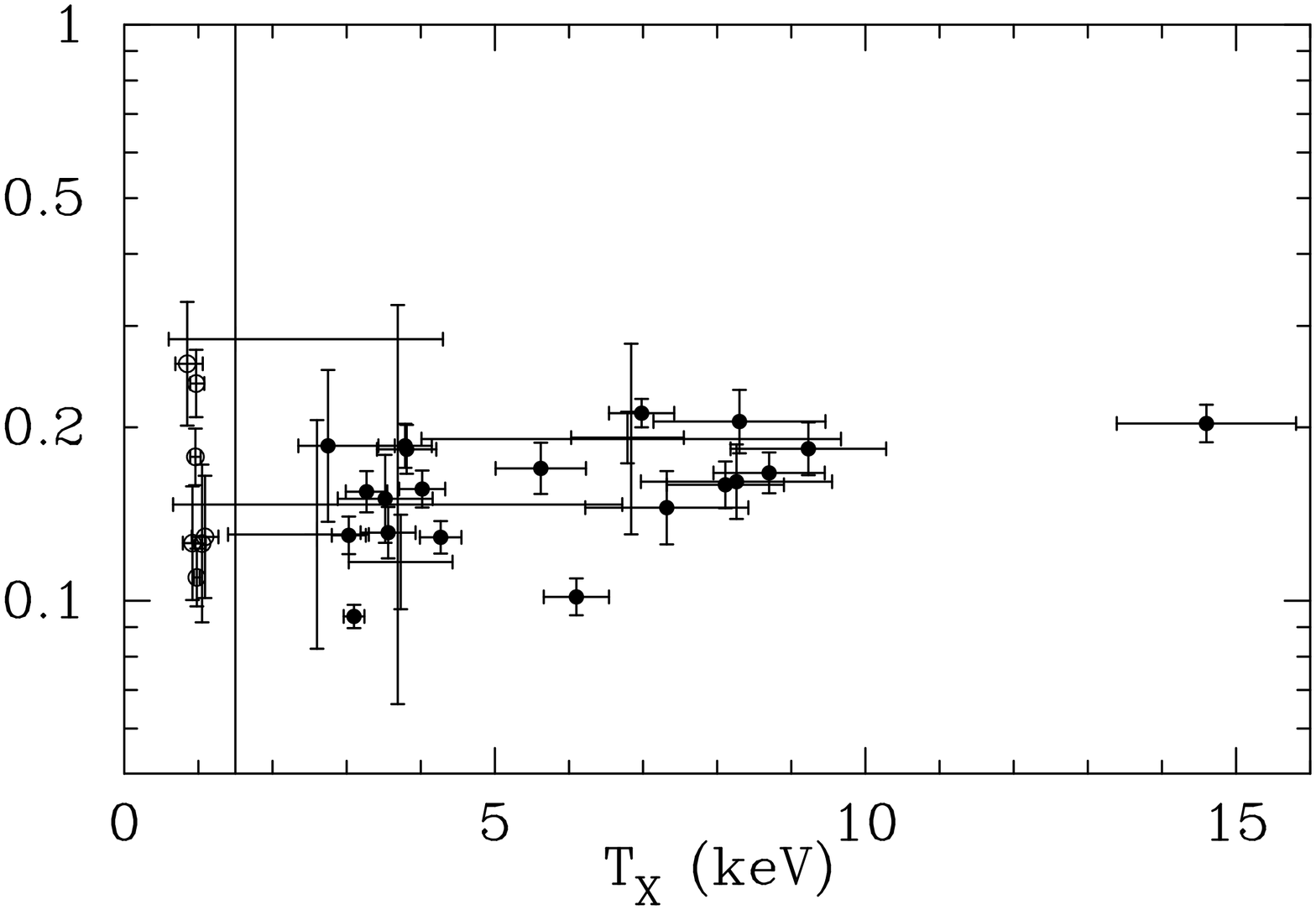}
\end{minipage}
\caption[]{Baryon fractions in the sample as a function of X-ray temperature. 
           Top~: at $\rxl$. Middle~: at $r_{2000}$. Bottom~: at $r_{200}$.
           Left~: hydrostatic masses. Right~: SLM masses. 
           Groups are shown as open circles and objects with poor quality
           temperature measurements (and therefore masses) as crosses.\label{fig:frac}}
\end{figure*}

In order to properly understand the baryon fraction in clusters it is necessary to understand what are the relative contribution of the gas and stellar components. Several previous studies found that the stellar component is more dominant in low temperature systems, the lower gas content of small clusters being possibly due to feedback processes. Our sample, being large and covering temperatures from 1 to 14 keV, allows us to study these questions in detail. 

\subsection{The $\mgas/\mst\:$--$\:\tx$ correlation}

\hspace*{1em}
    In this section, we examine a possible correlation between the X-ray gas temperature
and  the ratio of gas mass to stellar mass, $\mgas/\mst$, at various radii. A strong correlation has been previously found by D90: from the analysis of twelve groups and clusters with temperatures ranging
from 1 to $9 \kev$, D90 found that this ratio varies  by more than a factor of five from groups to rich clusters. An increase of $\mgas/\mst$
with cluster richness has also been reported by Arnaud et al. (1992). \\
This trend has been interpreted as due to the galaxy formation being less efficient in hot clusters than in colder systems. D90 suggest that the scenario for
structure formation of hierarchical clustering, in which large structures
form after little ones by successive mergers, is adequate to explain their
result~: in fact, as mergers go on, the intra-cluster gas is progressively
heated by shocks to higher and higher temperatures, as the size of the 
structures
 involved increases~; the higher $\tx$, the more difficult it becomes
for the gas to collapse and to form new galaxies. Hence, after some time, further
galaxy formation would be prevented in hot clusters, producing an anti-bias. \\
\hspace*{1em}
\hspace*{1em}
  In figure \ref{fig:mtg} we have plotted $\mgas/\mst$ against temperature at the 
radii $r_{2000}$ and $r_{200}$. As can been seen, 
mean figures for groups, cool clusters and hot clusters seem to show the sequence
observed by D90, but in a less pronounced  way: we find that cool clusters have $\mgas/\mst$ which is $\sim 3$ times smaller than for hotter ones instead of a factor of $\geq 5$ in D90. Moreover, this  apparent sequence weakens when plotted at $ r_{200}$: $\mgas/\mst$ is only twice smaller for groups than for hot clusters. 

It should be kept in mind that
in figure \ref{fig:mtg}, we adopted a constant galactic mass to
luminosity ratio for clusters and groups, whereas it is expected to be lower
for late type galaxies than for E-S0. As morphological segregation tends to
raise the fraction
of early type galaxies in rich clusters, taking into account this variation
of $\mst/L$ with morphological type would in fact flatten further the
observed correlation between $\mgas/\mst$ and $\tx$, as would do
taking into account the difference in galactic output from groups to clusters. We conclude that our sample does not show a strong evidence, if any, of increasing $\mgas/\mst$ with $\tx$ as previously found by D90.

\subsection{The $\fb\:$--$\:\tx$ correlation}

\hspace*{1em}
   Our sample spans an unprecedented wide range of temperature, allowing to test the somewhat 
puzzling evidence
that cool clusters have lower mean gas fraction than hot clusters. 
This trend has
been first reported by D95 and seems to be confirmed (Arnaud \& Evrard 1999). 
A modest increase of the gas fraction with $T_{X}$ has also been reported by 
Mohr et al. (1999). Such a trend is unexpected  in a self-similar 
cluster evolution,
$\fg$ and $\fb$ at a given overdensity
being expected to be constant, but  would be naturally explained by non gravitational processes such as galaxy feedback (for instance, early 
supernovae-driven
galactic outflows), able to heat the intergalactic gas enough to make it expand 
out (Metzler \& Evrard 1994, 1997; Ponman et al. 1999). This is achieved more easily in shallower 
potential wells
like those of groups, which could even experience substantial gas expulsion, 
thus reducing their gas
fractions. Such scenarii are necessary to explain the $L_X-T_X$ relation (Cavaliere et al., 1997).\\
In  order to examine this issue, we plot in figure \ref{fig:frac} the baryon fraction versus the temperature at different radii: $R_{Xlim}$, $r_{2000}$ and $r_{200}$. Error bars were estimated by considering  
uncertainties on the temperature and on metallicity for groups. Uncertainties on X-ray emission are small and leads to tiny uncertainties on the gas mass in 
the observed range ($R < R_x$), while in the outer part, where observations are 
lacking,  robust estimates of the uncertainties cannot be obtained, given that these uncertainties are 
systematic in nature. In the case of groups,  metallicity
      uncertainty can lead to significant uncertainties on gas mass, and was therefore taken into account. As it can be seen, we do observe no obvious trend with $\tx$. The data are more consistent with $\fb$ being constant and this whatever the mass estimator used. Although a weak tendency
could be seen (in the frame of SLM masses), it appears swamped in the high
dispersion affecting objects of a same temperature. Therefore we do not confirm the trend of increasing $f_b$ with $T_X$ (or size) as previously found  by D95. This is a rather robust conclusion as our sample covers a wide range of temperature, from 1 to 14 keV.
 This result is consistent with the similarity of the baryon fraction profiles we found (Fig. \ref{fig:tprofiles}) and the absence of trend of $\mgas/\mst$ with $\tx$ indicating that non-gravitational processes such as galactic feedback are not dominant in determining the large scale structure of  the intracluster medium. 
%Moreover, we find that inside a same object, the gas and baryon fraction increase
%from the center to the outer shells (Fig. \ref{fig:tprofiles} and Fig. %\ref{fig:mprofiles}), reflecting the fact that the gas seems to have a flatter %distribution than
%the dark matter a trend similar to what is found by D95.

\subsection{The $\fb\:$--$\:\beta$ correlation} 

\hspace*{1em}
Analysing the baryon fraction versus temperature may hide or reflect some correlation which are present among other parameters. Of special interest is to check whether a correlation with $\beta$ exists.
\begin{figure*}[!ht]
\hspace*{-0.5cm}
\begin{minipage}[t]{9cm}
\includegraphics[scale=0.35,angle=-90]{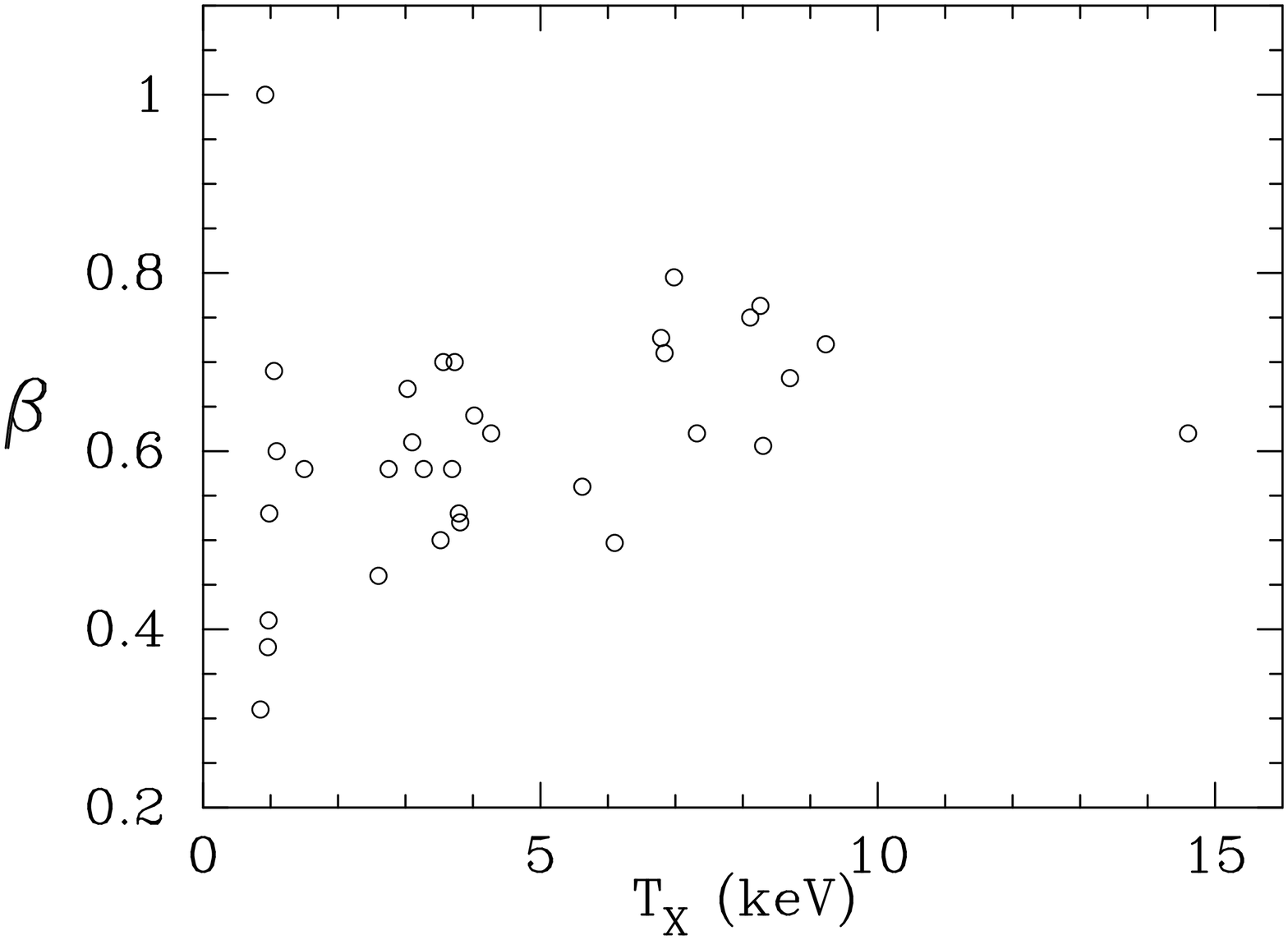}
\end{minipage}
\begin{minipage}[t]{9cm}
\includegraphics[scale=0.35,angle=-90]{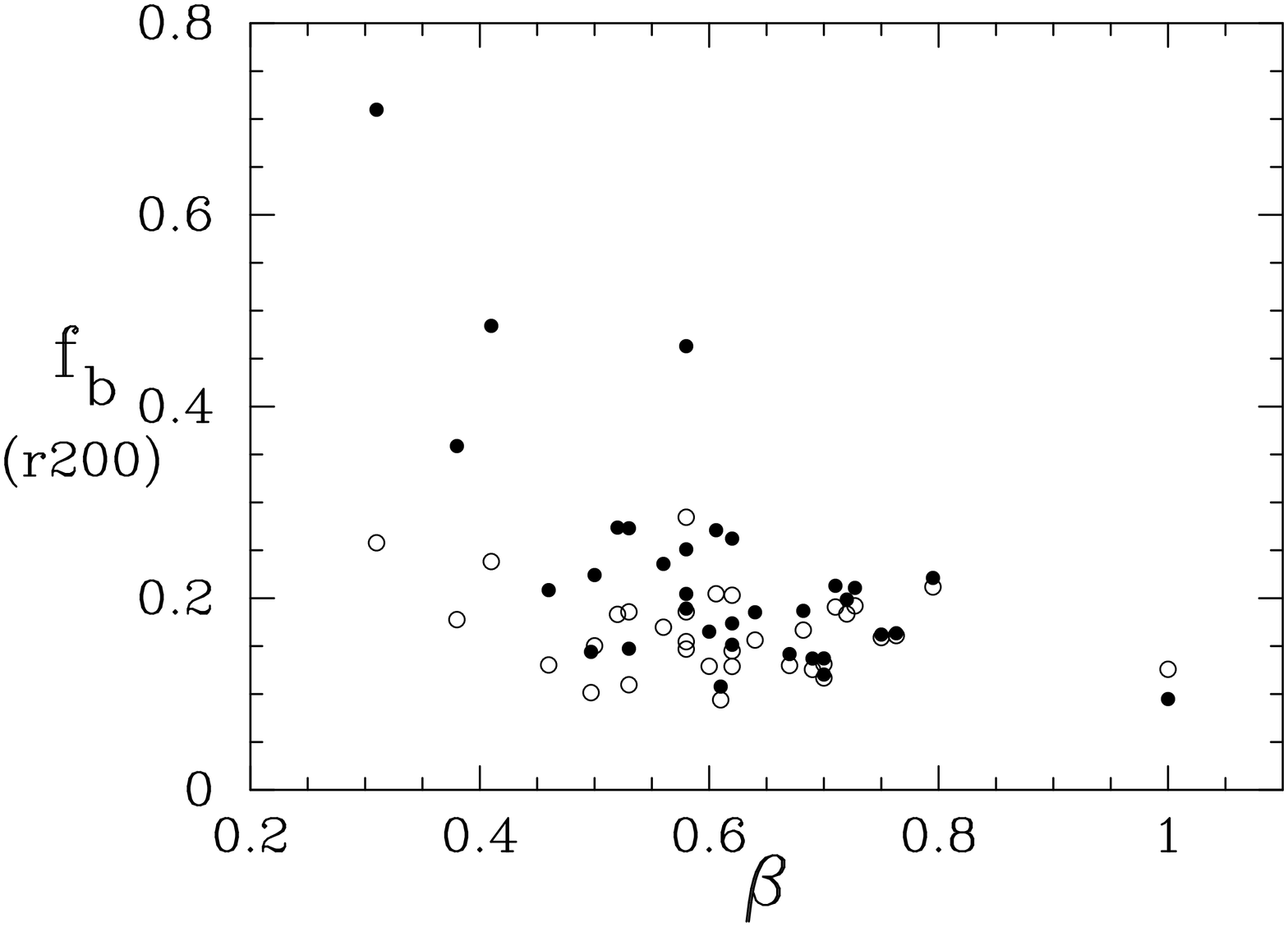}
\end{minipage}
\caption{The slope $\beta$ derived from the best-fit of a \bmodel
         to X-ray images plotted versus the temperature. On the
         {\it rigth panel} is plotted the baryon fraction (at $r_{200}$)
in the
         IHE \bmodel (filled symbols) compared to the SLM (open symbols)         as a function of $\beta$. \label{fig:betam}}
\end{figure*}

We first searched for a trend between $\beta$ and the temperature. Previous studies have shown that low temperature systems exhibit a more extended ICM distribution (low $\beta$ values) than hotter ones (Arnaud \& Evrard 1999). From figure \ref{fig:betam} we can see that no clear trend of increasing $\beta$ with $T_X$ is found. Although  
%the mean $\beta$ is 0.55 for groups ($T-x \sim 1$ keV), while it is closer to %0.7 for $T_x \geq 7$keV. 
smaller $\beta$ are found at the cool side, this might be  due to a larger dispersion in $\beta$ for the smallest potentials. We note that our result is consistent with the recent analysis of Mohr et al. (1999). 

We have also examined the way the baryon fraction varies with $\beta$ (Fig. \ref{fig:betam}). The baryon fraction derived from the hydrostatic $\beta$ model, $f_{b}^{IHE}$, does not vary with $\beta$ in an obvious way: if anything it decreases with increasing $\beta$ (while no obvious correlation is found with $\tx$, see Fig. \ref{fig:frac}).
Such a trend if real would be unexpected. Using the SLM mass estimates, the baryon fraction is much more constant and  less dispersed, even at a fixed  $\beta$ (for $\beta \sim 0.5-0.7$ the dispersion on $f_{b}/\overline{ f_{b}}$ is 0.23 with SLM estimates while it is 0.31 with the IHE). The fact that $f_{b}$ is constant with $\beta$ again differs from what one would expect if reheating would have a dominant role in redistributing the gas inside clusters.

\subsection{Implications on mass estimates}

\hspace*{1em}
As we have seen (section 4.2) the baryon fraction estimated with the SLM method is less dispersed than with the IHE method. 
%This fact implies that from a practical point of view the IHE estimator is less %robust than the SLM. 
This effect has been noticed previously (Evrard, 1997) and  has been interpreted as due to 
the observational uncertainties in the estimation of $\beta$. The mass estimates (at some radius $R$) can be written as:
\begin{equation}
M_1 = a_1 T R
\end{equation}
and for the IHE model~:
\begin{equation}
M_2 \sim a_2 \beta T R.
\end{equation}
The fact that baryon fractions estimated with $M_2$ are more dispersed
can be understood just because of the extra dispersion introduced by $\beta$ (EMN~; Arnaud \& 
Evrard, 1999).
For this to be due to the sole errors in the measurement of 
$\beta$, it would imply that the dispersion in the measurements dominates
the intrinsic dispersion, resulting in a tight correlation between $\beta$ and $\fb$, which is not obvious from Fig. \ref{fig:betam}: 
most clusters have a $\beta$ in the narrow range $0.5-0.7$, and the sample restricted to this range shows a larger dispersion for the
baryon fraction computed with the IHE. Therefore, we conclude that the large dispersion observed in the baryon fractions estimated from the hydrostatic $\beta$ model is intrinsic to the method itself leading to less reliable mass estimates, rather than to the uncertainty on $\beta$ measurements.
 
\section{Discussion and conclusions}

\hspace*{1em}
   We have analysed a sample of 33 galaxy clusters and groups covering a wide 
range of temperatures. For all clusters, X-ray and optical data were gathered 
from the literature
(except for Abell 665 whose ROSAT PSPC data have been reanalysed by us). 
This has allowed us to investigate the structure of the various baryonic components 
of X-ray clusters. Mass estimates were derived from two different methods~:
first we have followed the standard hydrostatic isothermal equation (IHE method), secondly 
we have estimated the virial mass and mass profile
by using the universal dark matter profile of NFW in which the virial radius
is deduced from the scaling relation argument (SLM method), the normalisation constant being taken from EMN and from BN. \\
\hspace*{1em}
   We find that virial 
masses (i.e. masses enclosed inside a fixed contrast density radius) are 
systematically and significantly lower when one is using the hydrostatic 
isothermal equation.  After this paper was submitted, we have been aware of a 
recent similar study by Nevalainen et al. (2000) who found that taking into account temperature profiles exacerbates this difference, as inferred masses are then smaller. Examination of the baryon fraction versus contrast 
density 
has shown that the baryon fraction is more dispersed using the IHE. We have shown that this is not due to uncertainties on the $\beta$ measurement but rather reflects the fact that the IHE method does not provide as reliable a mass estimate as the SLM, neither in the inner parts nor in the outer regions.
 Moreover the tightening of $f_{b}^{SLM}$ profiles supports the idea that baryon profiles in clusters 
do have a rather regular structure, i.e. that gas distribution is nearly self-similar
 which is consistent with the recent studies  by Vikhlinin et al.(1999) and by Neumann \& Arnaud (1999) who found an evidence of regularity in gas density 
profiles. However, when plugging their mean standard density 
profile into the hydrostatic equation, these last authors found a mean total mass profile 
which is different from the NFW profile (their 
mass profile is lower than the one derived from numerical 
simulations). They have used the hydrostatic 
isothermal equation to estimate their total mass which probably explains this discrepancy.\\
Our mean gas fraction at the virial radius $r_{200}$, using the SLM, is found to be in the range 12.6--14.6 $\%$
(for $h = 0.5$), in rough agreement with Arnaud and Evrard (1999) when EMN normalization is used. The 
mean baryonic fraction is $f_{b}^{SLM} \approx 13.7-16.4$.
It is important to emphasize that our analysis shows that a 
larger baryon fraction could
be obtained when the sole hydrostatic equation is used, but it is reasonable 
to think that this is 
an overestimation due to the mass estimator itself.  Our analysis is consistent with an intrinsic dispersion of 20\% in baryon fractions (but this could be due to some systematics), which means that our mean baryon fraction is uncertain by less than 0.01. \\

As the observed luminosity-temperature does not follow the simple scaling expected from self-similarity, it is likely that  non--gravitational heating such as galactic winds, additional energy input by Type II supernovae, play an important role in the physics of the X-ray gas and may result in  inflating the gas distribution. Metzler \& Evrard (1997) have studied this possibility and found that this is achieved more easily in low temperature clusters (shallow potential wells). The consequence of such an effect is an increasing gas fraction outwards within a cluster and a decreasing gas fraction with decreasing temperature. This effect is expected to be more pronounced in groups and cold clusters. From our sample, we do not observe such a trend of the baryon (gas) fraction with $\tx$ whatever the method we use, suggesting that non gravitational heating is not playing a dominant role on the scale of the virial radius. In turn, we confirm that the baryon fraction apparently increases significantly from the center to outer parts of clusters. \\
Several previous studies have shown a clear trend of increasing $\beta$ with $\tx$, while we do not find such a clear trend, neither we find a trend of $f_{b}$ with $\beta$ which is consistent with the absence of an $f_{b}$--$ {\tx}$ correlation, although we find a slightly decreasing $f_{b}^{IHE}$ (derived from the IHE method) with increasing $\beta$, but with a large dispersion. This result is important as it shows that the scatter in the baryon fractions derived from the IHE method is probably not due to the sole errors in the $\beta$ measurement, but is rather due to the IHE mass estimator itself. \\
Our sample does not show any evidence of the strong indication highlighted 
by D90 that in low temperature systems, a larger fraction 
of baryons is present in the stellar component. Although a trend could be 
present in our sample, the data are certainly consistent with a stellar to gas 
mass ratio being constant with temperature (or mass), a further argument that non gravitational processes such as galactic winds are playing a minor role in the overall distribution of gas in clusters. \\
\hspace*{1em}
Finally, it appears that the properties of X-ray clusters are still difficult 
to quantify because of the lack of large homogenous samples of clusters 
for which both optical and X-ray data are available. Such a situation is 
likely to improve with Chandra and XMM. Nevertheless, the sample 
we have studied reveals that clusters show important differences in 
the detail of the structure of their baryonic content, but that their global 
properties,
baryonic fraction and stellar content do not show strong systematic differences
with temperature. \\
\vspace*{2ex}

\noindent
\textbf{Notes on individual clusters :}

\begin{list}{--}{}
\item A85 : Optical data for this cluster are unsafe and extend only out to 900 kpc.
            We used the observations of Murphy (1984) but, as the fit with an
            unusual galaxy density profile he performs is rather poor (and
            suffers from an inconsistency between $H_0 = 50$ and $H_0 = 60$), we
            chose to replace it with a standard King profile.
\item A401 : As Buote \& Canizares (1996) do not give the central electron
             density, we computed it with our program in the same way as for
             Abell 665, since both objects were observed with ROSAT PSPC, using
             the galactic hydrogen column density of David et al. (1993)
             and the count rate inside a given radius provided by Ebeling 
             et al. (1996).
\item A2029 : The same as for A401 applies.
\item A2163 : Optical data for this cluster are unsafe.
              No galaxy distribution was available. We therefore fitted the
              integrated luminosity profile given in Squires et al. (1997),
              but it was not corrected for background galaxies
              and it extends only to $1.3 \mpc$ whereas $\rxl = 4.6 \mpc$.
\item AWM7 : We used the list of galactic positions and magnitudes within $1^o$
             of the central cD of Beers et al. (1984) to build an integrated
             luminosity profile, corrected for incompleteness using their
             limiting magnitude and the standard Schechter luminosity function.
             The optical core radius was imposed to be the same as the X-ray core
             radius, which gives very similar results as excluding the three
             innermost galaxies (otherwise, the fitted core radius is too small
             and in fact, the King form is not a good representation of the
             central parts of clusters).
\item Hydra A : The same procedure as for A2163 was applied, with the three points
                given by D90~:
                $L_{\rm V} / L_{\rm V \odot} = 8.2\:10^{11}$ at $0.5 \mpc$,
                $1.3\:10^{12}$ at $1 \mpc$ and $1.9\:10^{12}$ at $2 \mpc$.
\item HCG 62 : We derived an integrated number count profile from the list of
               galactic positions of Zabludoff \& Mulchaey (1998) (using their
               velocity criteria to select true members) and fitted it with
               the function in Equ.7, assigning to each galaxy the mean
               luminosity derived from the limiting magnitude of the observations
               and the standard Schechter luminosity function.
               The number of galaxies contained in this ``compact group''
               is much larger than usually assumed (45 members with
               $\mbapp < 17$ instead of 4 in Hickson 1982).
\item HCG 94 : Ebeling et al. (1995) claim this object has been
               misclassified and, from its X-ray emission, looks more like a
               poor cluster rather than a compact group. Only 7 galaxies
               are generally attributed to HCG 94 but we made use
               of the indication of Ebeling et al. that 12 more galaxies
               are observed within a $1 \mpc$ radius and at $\mbapp \leq 18$,
               to which we attribute a mean luminosity as for HCG 62.
               The fit by Equ.7 we perform relies entirely on this point
               (fixing the core radius at the X-ray value) since inclusion of
               the central galaxies would lead to a physically unacceptable core
               radius). Therefore, optical data for this object are unsafe.
\item NGC 533 : As no central electron density is given by Mulchaey et al. (1996),
                we computed it from the gas mass that they obtain at a given distance.
                For the optical part, this is the same case as HCG 62.
                This group contains 36 members with $\mbapp < 17$ instead of
                4 in Geller \& Huchra (1983).
\item NGC 2300 : The same as for AWM7 applies, using magnitudes from the RC3
                 and excluding the two central galaxies from the fit instead of
                 fixing the core radius.
\item NGC 4261 : Since the central electron density Davis et al. (1995)
                 give is inconsistent with their total gas mass, we computed it
                 (this is again the same case as for A401 and A2029). We also used
                 optical data directly from Nolthenius (1993) and applied
                 the same method as for AWM7 with magnitudes taken from the RC3
                 (except we did not have to impose the core radius).
\item RXJ 1340.6 has not been included in figures showing mass ratios as a
      function of overdensity, because it is a very peculiar case~:
      the interior of the central giant elliptical galaxy is seen through
      a very large range of overdensities (at least out to $\delta = 7000$).
\item Several clusters have unreliable X-ray temperatures~:
      A76, A426 (very strong cooling flow), A1377, A1775
      (likely very strong cooling flow) and A2218 (steeply outwards-decreasing
      temperature profile).
\item Error bars on gas mass for groups include an estimate of the metallicity
      uncertainty, which results in an uncertainty on the electron density.
      This effect was taken into account only for groups, because it is
      significant mostly in the case of low temperatures.

\end{list}

\begin{acknowledgements}We acknowledge useful discussions with M. Arnaud, D. Neumann 
and J. Bartlett, and we thank V. Pislar for much appreciated help in ROSAT PSPC
data reduction.
\end{acknowledgements}


\begin{thebibliography}{}
\bibitem {}Arnaud M. \& Evrard A.\ E., 1999, MNRAS 305, 631
\bibitem {}Arnaud M., Rothenflug R., Boulade O., Vigroux L. \& Vangioni-Flam E., 1992, A\&A 254, 49
\bibitem {}Balland C. \& Blanchard A., 1997, ApJ 487, 33
\bibitem {}Bartlett J.\ G., 1997, ASP Conf. Series 126, 365
\bibitem {}Beers T.\ C., Geller M.\ J., Huchra J.\ P., Latham D.\ W. \& Davis R.\ J., 1984, ApJ 283, 33 (B84)
\bibitem {}Briel U.\ G., Henry J.\ P. \& B\"ohringer H., 1992, A\&A 259, L31 (BHB92)
\bibitem {}Bryan G.\ L. \& Norman M.\ L., 1998, ApJ 495, 80 (BN)
\bibitem {}Buote D.\ A. \& Canizares C.\ R., 1996, ApJ 457, 565 (BC96)
\bibitem {}Butcher H. \& Oemler A., 1978, ApJ 226, 559 (BO78)
\bibitem {}Cavaliere A., Menci N. \& Tozzi P., 1997, ApJ 484, L21
\bibitem {}Cavaliere A. \& Fusco-Femiano R., 1976, A\&A 49, 137
\bibitem {}Cirimele G., Nesci R. \& Tr\`evese D., 1997, ApJ 475, 11 (CNT97)
\bibitem {}Cruddace R.\ G., Kowalski M.\ P., Fritz G.\ G. et al., 1997, ApJ 476, 479 (CK97)
\bibitem {}David L.\ P., Jones C. \& Forman W., 1995, ApJ 445, 578 (D95)
\bibitem {}David L.\ P., Jones C., Forman W. \& Daines S., 1994, ApJ 428, 544 (D94)
\bibitem {}David L.\ P., Slyz A., Jones C. et al., 1993, ApJ 412, 479 (D93)
\bibitem {}David L.\ P., Arnaud K.\ A., Forman W. \& Jones C., 1990, ApJ 356, 32 (D90)
\bibitem {}Davis D.\ S., Mulchaey J.\ S., Mushotzky R.\ F. \& Burstein D., 1996, ApJ 460, 601 (DM96)
\bibitem {}Davis D.\ S., Mushotzky R.\ F., Mulchaey J.\ S. et al., 1995, ApJ 444, 582 (DM95)
\bibitem {}Dell'Antonio I.\ P., Geller M.\ J. \& Fabricant D.\ G., 1995, AJ 110, 502
\bibitem {}Dressler A., 1978, ApJ 223, 765 (D78a)
\bibitem {}Dressler A., 1978, ApJ 226, 55 (D78b)
\bibitem {}Durret F., Gerbal D., Lachi\`eze-Rey M., Lima-Neto G. \& Sadat R., 1994, A\&A 287, 733
\bibitem {}Ebeling H., Voges W., B\"ohringer H. et al., 1996, MNRAS 281, 799 (EVB96)
\bibitem {}Ebeling H., Mendes de Oliveira C. \& White D.\ A., 1995, MNRAS 277, 1006 (EMW95)
\bibitem {}Elbaz D., Arnaud M. \& B\"ohringer H., 1995, A\&A 293, 337 (EAB95)
\bibitem {}Evrard A.\ E., 1997, MNRAS 292, 289
\bibitem {}Evrard A.\ E., Metzler C.\ A. \& Navarro J.\ F., 1996, ApJ 469, 494 (EMN)
\bibitem {}Ferguson H.\ C. \& Sandage A., 1990, AJ 100, 1 (FS90)
\bibitem {}Ferguson H.\ C., 1989, AJ 98, 367 (F89)
\bibitem {}Ferguson H.\ C. \& Sandage A., 1988, AJ 96, 1520 (FS88)
\bibitem {}Frenk C.\ S., White S.\ D.\ M. et al. 1999, ApJ 525, 554
\bibitem {}Fukazawa Y., Makishima K., Tamura T. et al., 1998, PASJ 50, 187 (F98)
\bibitem {}Garcia A.\ M., 1993, A\&AS 100, 47 (G93)
\bibitem {}Geller M.\ J. \& Huchra J.\ P., 1983, ApJS 52, 61
\bibitem {}Hammer F., 1991, ApJ 383, 66
\bibitem {}Henry J.\ P., Briel U.\ G. \& Nulsen P.\ E.\ J., 1993, A\&A 271, 413 (HBN93)
\bibitem {}Hickson P., Kindl E. \& Auman J.R., 1989, ApJS 70, 687 (HKA89)
\bibitem {}Hickson P., 1982, ApJ 255, 382
\bibitem {}Hughes J.\ P. \& Tanaka Y., 1992, ApJ 398, 62
\bibitem {}Ikebe Y., Ezawa H., Fukazawa Y. et al., 1996, Nature 379, 427 (I96)
\bibitem[Irwin, Bregman and Evrard (1999)]{1999ApJ...519..518I} Irwin, J.\ 
A., Bregman, J.\ N.\ \& Evrard, A.\ E.\ 1999, ApJ 519, 518 
\bibitem {}Kent S.\ M. \& Sargent W.\ L.\ W., 1983, AJ 88, 697 (KS83)
\bibitem {}Kent S.\ M. \& Gunn J.\ E., 1982, AJ 87, 945 (KG92)
\bibitem {}Loewenstein M. \& Mushotzky R.\ F., 1996, ApJ 471, L83 (LM96)
\bibitem {}Lugger P.\ M., 1989, ApJ 343, 572
\bibitem[Markevitch, Forman, Sarazin and Vikhlinin 
(1998)]{1998ApJ...503...77M} Markevitch M., Forman W.\ R., Sarazin C.\ L. 
\& Vikhlinin A., 1998, ApJ, 503, 77 (MF98)
\bibitem {}Markevitch M., 1998, ApJ, 504, 27
\bibitem {}Metzler C.\ A. \& Evrard A.\ E., 1997, ApJ submitted, astro-ph/9710324
\bibitem {}Metzler C.\ A. \& Evrard A.\ E., 1994, ApJ 437, 564
\bibitem {}Mewe R., Lemen J.\ R. \& van den Oord G.\ H.\ J., 1986, A\&AS 65, 511
\bibitem {}Mohr J.\ J., Mathiesen B. \& Evrard A.\ E., 1999, ApJ 517, 627
\bibitem {}Mohr J.\ J., Geller M.\ J., Fabricant D.\ G. et al., 1996, ApJ 470, 724 (MG96)
\bibitem {}Morrisson R. \& McCammon D., 1983, ApJ 270, 119
\bibitem {}Mulchaey J.\ S., Davis D.\ S., Mushotzky R.\ F. \& Burstein D., 1996, ApJ 456, 80 (MD96)
\bibitem {}Murphy H.\ P., 1984, MNRAS 211, 637 (M84)
\bibitem {}Navarro J.\ F., Frenk C.\ S. \& White S.\ D.\ M., 1995, MNRAS 275, 720 (NFW)
\bibitem {}Neumann D.\ M. \& Arnaud M., 1999, A\&A 348, 711
\bibitem {}Neumann D.\ M. \& B\"ohringer H., 1995, A\&A 301, 865 (NB95)
\bibitem[Nevalainen, Markevitch and Forman (2000)]{2000ApJ...532..694N} 
Nevalainen J., Markevitch M. \& Forman W., 2000, ApJ 532, 694 
\bibitem {}Nolthenius R., 1993, ApJS 85, 1 (N93)
\bibitem {}Oegerle W.\ R. \& Hoessel J.\ G., 1989, AJ 98, 1523 (OH89)
\bibitem {}Oegerle W.\ R., Hoessel J.\ G. \& Jewison M.\ S., 1987, AJ 93, 519 (OHJ87)
\bibitem {}Oukbir J. \& Blanchard A., 1992, A\&A 262, L210
\bibitem {}Pen U.-L., 1998, ApJ, 498, 60
\bibitem {}Pislar V., Durret F., Gerbal D., Lima Neto G.\ B. \& Slezak E., 1997, A\&A 322, 53 (P97)
\bibitem {}Ponman T.\ J., Cannon D.\ B. \& Navarro J.\ F., 1999, Nature 397, 135
\bibitem {}Ponman T.\ J., Allan D.\ J., Jones L.\ R. et al., 1994, Nature 369, 462
\bibitem {}Ponman T.\ J. \& Bertram D., 1993, Nature 363, 51 (PB93)
\bibitem {}Rhee G.\ F. \& Latour H.\ J., 1991, A\&A 243, 38
\bibitem {}Schechter P., 1975, Ph.D. thesis
\bibitem {}Schneider D.\ P., Gunn J.\ E. \& Hoessel J.\ G., 1983, ApJ 264, 337
\bibitem {}Snowden S.\ L., McCammon D., Burrows D.\ N. \& Mendenhall J.\ A., 1994, ApJ 424, 714
\bibitem {}Squires G., Neumann D.\ M., Kaiser N. et al., 1997, ApJ 482, 648 (SN97)
\bibitem {}Squires G., Kaiser N., Babul A. et al., 1996, ApJ 461, 572 (SK96)
\bibitem {}Tamura T., Day C.\ S., Fukazawa Y. et al., 1996, PASJ 48, 671 (T96)
\bibitem {}Tyson J.\ A., Valdes F. \& Wenk R.\ A., 1990, ApJ 349, L1
\bibitem {}van der Marel R.\ P., 1991, MNRAS 253, 710
\bibitem {} Vikhlinin A., Forman W. \& Jones C. 1999, ApJ, 525, 47
\bibitem {}White D.\ A., 2000, MNRAS 312, 663 (W00)
\bibitem {}White D.\ A., Jones C. \& Forman W., 1997, MNRAS 292, 419 (WJF97)
\bibitem {}White D.\ A. \& Fabian A.\ C., 1995, MNRAS 273, 72
\bibitem {}White S.\ D.\ M., Navarro J.\ F. \& Evrard A.\ E., 1993, Nature 366, 429
\bibitem {}Zabludoff A.\ I. \& Mulchaey J.\ S., 1998, ApJ 496, 39 (ZM98)
\bibitem {}Zwicky F., 1933, Helv. Phys. Acta 6, 110
\end{thebibliography}
\end{document}